\documentclass[aps,twocolumn,showpacs,pra,superscriptaddress,floatfix,longbibliography]{revtex4-1}

\usepackage{amsmath,amsthm,amsfonts,amssymb,times,bbm,graphicx,longtable}
\usepackage{color,tabularx,epsfig,wasysym,dcolumn,bm,epstopdf,subfigure}
\usepackage{bbold}

\newcommand{\bd}[1]{\boldsymbol{#1}}
\newcommand{\bra}[1]{\mbox{$\langle #1 |$}}
\newcommand{\ket}[1]{\mbox{$| #1 \rangle$}}

\newcommand{\N}{{\mathbb{N}}}
\newcommand{\R}{{\mathbb{R}}}

\begin{document}
\title{Influence of the Fermionic Exchange Symmetry beyond Pauli's Exclusion Principle}

\author{Felix Tennie}
\affiliation{Clarendon Laboratory, University of Oxford, Parks Road, Oxford OX1 3PU, United Kingdom}
\author{Vlatko Vedral}
\affiliation{Clarendon Laboratory, University of Oxford, Parks Road, Oxford OX1 3PU, United Kingdom}
\affiliation{Centre for Quantum Technologies, National University of Singapore, 3 Science Drive 2, Singapore 117543}
\author{Christian Schilling}
\email{christian.schilling@physics.ox.ac.uk}
\affiliation{Clarendon Laboratory, University of Oxford, Parks Road, Oxford OX1 3PU, United Kingdom}

\date{\today}

\begin{abstract}
Pauli's exclusion principle has a strong impact on the properties of most fermionic quantum systems. Remarkably, the fermionic exchange symmetry implies further constraints on the one-particle picture. By exploiting those generalized Pauli constraints we derive a measure which quantifies the influence of the exchange symmetry beyond Pauli's exclusion principle. It is based on a geometric hierarchy induced by the exclusion principle constraints. We provide a proof of principle by applying our measure to a simple model. In that way, we conclusively confirm the physical relevance of the generalized Pauli constraints and show that the fermionic exchange symmetry can have an influence on the one-particle picture beyond Pauli's exclusion principle. Our findings provide a new perspective on fermionic multipartite correlation since our measure allows one to distinguish between static and dynamic correlations.
\end{abstract}


\maketitle
\section{Introduction}\label{sec:intro}
The properties and the behavior of most fermionic quantum systems strongly rely on Pauli's famous exclusion principle \cite{Pauli1925}.
This principle defines a constraint on the one-particle picture: For any $N$-fermion state \ket{\Psi_N} the occupancies of one-particle states $\ket{\varphi}$  are restricted, $0\leq \bra{\Psi_N} \hat{n}_{\varphi}\ket{\Psi_N}\leq 1$. Indeed, this constrains the one-particle reduced density matrix (1RDM) $\rho_1\equiv N tr_{N-1}[\ket{\Psi_N}\bra{\Psi_N}]$, obtained by tracing out $N-1$ fermion, according to
\begin{equation}\label{PCrho}
\mathbb{0}\leq \rho_1\leq \mathbb{1}\,.
\end{equation}
Equivalent to this matrix relation, the natural occupation numbers $\lambda_i$, i.e.~the eigenvalues of the 1RDM, are restricted,
\begin{equation}\label{PC}
0 \leq \lambda_i \leq 1\,.
\end{equation}
These Pauli constraints (PC) play an important role for various physical phenomena with remarkable consequences for both, the micro and the macro world. On a microscopic scale they are the basis of the `Aufbau principle' for atoms and nuclei. For macroscopic systems the Pauli exclusion principle is responsible for the stability of matter \cite{Dyson1967,LiebStab}. In particular, for an ideal gas of fermions it implies an effective pressure, the Fermi degeneracy pressure, explaining on a rudimentary level the stability of neutron stars. Further emerging phenomena are the Fermi hole and the Pauli spin blockade. They originate from the fact that electrons with parallel spins are forbidden to sit at the same position.

This universal relevance of Pauli's exclusion principle is obvious for weakly correlated systems: All PC are (approximately) saturated, i.e.~one observes for each occupation number either $\lambda_i \approx 1$ or $\lambda_i \approx 0$. Such (approximate) \emph{pinning} by all PC is the typical behavior within the Landau-Fermi theory. Even for strongly correlated systems one observes this quasipinning by PC since at least the largest occupation numbers are very close to one and the smallest ones are very close to zero.

Despite the success on its own, the PC in the one-particle picture are \textit{de facto} a consequence of the antisymmetry of the $N$-fermion wavefunction \cite{Dirac1926,Heis1926}. This exchange symmetry is a much stronger requirement than Pauli's exclusion principle concerning the one-particle picture, only. It has been crucial, e.g., for the heuristic approach to the fractional quantum Hall effect. Using a Jastrow type wave function involving a product of antisymmetric two-particle functions $(z_i-z_j)^p$ with $p$ an odd, positive integer, Laughlin succeeded to explain a huge class of plateaus with fractional Hall conductivity  \cite{Laughlin1983}.

Since the exchange symmetry concerns the $N$-particle picture an important question arises: Does the fermionic exchange symmetry have an influence on one-particle properties \emph{beyond} the Pauli exclusion principle?
Until a few years ago it has even been unclear how to address this question in a meaningful way for concrete systems.
This has changed thanks to a recent mathematical breakthrough. Motivated by results for a very few special cases \cite{Borl1972}, it was shown that the fermionic exchange symmetry \emph{in general} implies further restrictions on the one-particle picture \cite{Kly3,Kly2,Altun}, completing Pauli's incomplete exclusion principle. Those generalized Pauli constraints will serve us to eventually address in the present work the long-standing question above. The corresponding findings exploiting an elegant quantum information theoretical language will provide a new perspective on fermionic multipartite correlation since they allow the distinction between static and dynamic correlation.

\section{Generalized Pauli constraints}\label{sec:GPC}
The so-called\emph{ generalized Pauli constraints} (GPC) take the form of linear inequalities
\begin{equation}\label{eq:gpc}
D_j(\vec{\lambda}) \equiv \kappa_j^{(0)}+\sum_{i=1}^d\kappa_j^{(i)} \lambda_i\geq 0\,,
\end{equation}
for the natural occupation numbers (NON) $\lambda_i$. These are the (decreasingly ordered)  eigenvalues of the 1RDM corresponding to the $N$-fermion state \ket{\Psi_N}. Here $\kappa_j^{(i)} \in \mathbb{Z}$, $j=1,2,\ldots,\nu^{(N,d)}<\infty$ and $d$ is the dimension of the underlying one-particle Hilbert space. Geometrically, for each fixed pair $(N,d)$ the family of GPC, together with the normalization and the ordering constraints $\lambda_1\geq\ldots\geq \lambda_d\geq 0$, defines a $(d-1)$-dimensional polytope $\mathcal{P}^{(N,d)}$ of allowed vectors $\vec{\lambda}\equiv (\lambda_i)_{i=1}^d$.

Each of the $D_j(\vec{\lambda})$ determines a $(d-2)$-dimensional facet $\mathcal{F}_j^{(N,d)}$ of $\mathcal{P}^{(N,d)}$,
\begin{equation}\label{eq:facet}
\mathcal{F}_j^{(N,d)} \equiv \{\vec{\lambda} \in \mathcal{P}^{(N,d)}\,|\,D_j(\vec{\lambda})=0\} \ .
\end{equation}
The remaining facets of $\mathcal{P}^{(N,d)}$ are defined by the saturation of the `ordering constraints', i.e.~by $\lambda_i =  \lambda_{i+1}$ for \mbox{$i=1,...,d-1$}.

To compare the GPC with the PC, notice that the PC (\ref{eq:PEPs}) alone define a $d$-dimensional `Pauli hypercube' $\mathcal{C}^{(d)}=[0,1]^d$. Accounting for the ordering and normalization of the NON, $\vec{\lambda}$ is then restricted to the \textit{Pauli simplex}
\begin{equation}\label{eq:PauliSimp}
\Sigma^{(N,d)} \equiv \{\vec{\lambda} \,|\,1\geq \lambda_1\geq \ldots\geq \lambda_d \geq 0 \,\,, \sum_{i=1}^d \lambda_i=N \}\ ,
\end{equation}
which is a $(d-1)$-dimensional subset of the hypercube $\mathcal{C}^{(d)}$.
It contains the polytope $\mathcal{P}^{(N,d)}$ as a proper subset confirming the incompleteness of Pauli's exclusion principle. For an illustration see Fig.~\ref{fig:polytope}. In the following we will typically skip the indices `$N$' and `$d$'.
\\

\section{The pinning phenomenon}\label{sec:pinning}
The potential influence of the fermionic exchange symmetry \emph{beyond} Pauli's exclusion principle is strongly linked to the \textit{boundary} of the polytope corresponding to the saturation of a GPC. First, it had been argued \cite{Kly1,Kly5} that  the ground state minimization process of the energy expectation value $\bra{\Psi}\hat{H}\ket{\Psi}$ for a model Hamiltonian $\hat{H}$ may get stuck on the boundary of the polytope $\mathcal{P}$, since any further minimization would violate some GPC.
This \emph{pinning} effect by GPC is physically relevant because it potentially restricts the dynamics of the corresponding system whose NON $\vec{\lambda}$ can never leave the polytope \cite{CSQMath12,CS2015Hubbard}.
This is in close analogy to the implications of the PC which forbid fermions at low temperature to decay to lower lying occupied energy levels. Second, pinning as an effect in the one-particle picture allows one to reconstruct the structure of the corresponding $N$-fermion quantum state which in addition is significantly simplified.

Let us explain the important latter point. For given $(N,d)$ an arbitrary $N$-fermion state $\ket{\Psi}$ can be expressed as a superposition of $\binom{d}{N}$ Slater determinants $\ket{\bd{i}}\equiv a_{i_1}^\dagger \ldots a_{i_N}^\dagger \ket{0}$, built from its own natural orbitals $\ket{i_{\nu}}$, the eigenstates of the corresponding 1RDM.
Now, let us assume that given $\vec{\lambda}$ is pinned by a GPC $D_j$, i.e.~$D_j(\vec{\lambda})=0$.
Then, the  corresponding superposition  is reduced to a subset  ${\mathcal{I}_{D_{j}}}$ containing only those configurations $\bd{i}$
for which $\hat{D_j}\ket{\bd{i}} = 0$  \cite{Kly1}.  $\hat{D_j}$ follows from  $D_j$ (Eq.~(\ref{eq:gpc})) by replacement of $\lambda_i$ by the occupation operator $\hat{n}_i$ of the natural orbital $\ket{i}$.
Accordingly, the general superposition of $\binom{d}{N}$ Slater determinants is reduced to (see also right side of Fig.~\ref{fig:SR2})
\begin{equation}\label{eq:PinPsi}
\ket{\Psi^{(pin)}} = \sum_{\bd{i}\in\mathcal{I}_{D_{j}}}\,c_{\bd{i}}\,\ket{\bd{i}}\ .
\end{equation}
This is quite similar to pinning by a PC, i.e.~$\lambda_i=1$ or $\lambda_i=0$. In that case, only those Slater determinants  $\ket{\bd{i}}$ contribute to $\ket{\Psi}$ which all \emph{do} or all \emph{do not} contain $i$, respectively.

Analytical \cite{CS2013,CS2015Hubbard,CS2016a,CS2016b} and numerical investigations \cite{BenavLiQuasi,Mazz14,RDMFT,BenavQuasi2,Mazz16,MazzGPC2RDM} have shown that pinning of the corresponding ground state NON  $\vec{\lambda}$  to the polytope's boundary does not occur, in general. Whereas pinning has been found for some model systems \cite{Mazz14,BenavQuasi2,CS2015Hubbard,RDMFT} other systems exhibit quasipinning, only \cite{CS2013,BenavLiQuasi,Mazz14,RDMFT,CS2016a}. In the latter case the vector of NON $\vec{\lambda}$ is close to but not exactly  on the polytope's boundary and various implications of pinning hold at least approximately \cite{CSQuasipinning}.
\\

\section{Hierarchy of pinning}\label{sec:hierarchy}
\begin{figure}[]
\centering
\includegraphics[width=4.3cm]{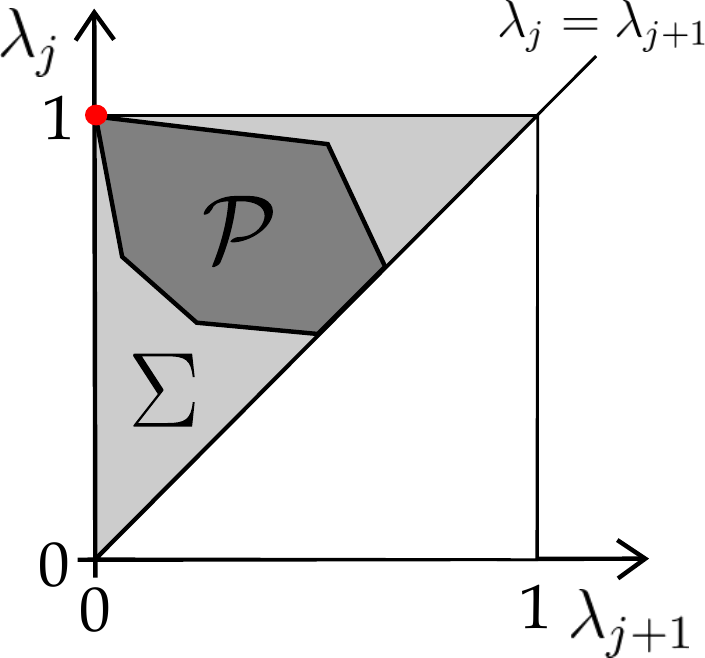}
\hspace{0.6cm}
\includegraphics[width=3.6cm]{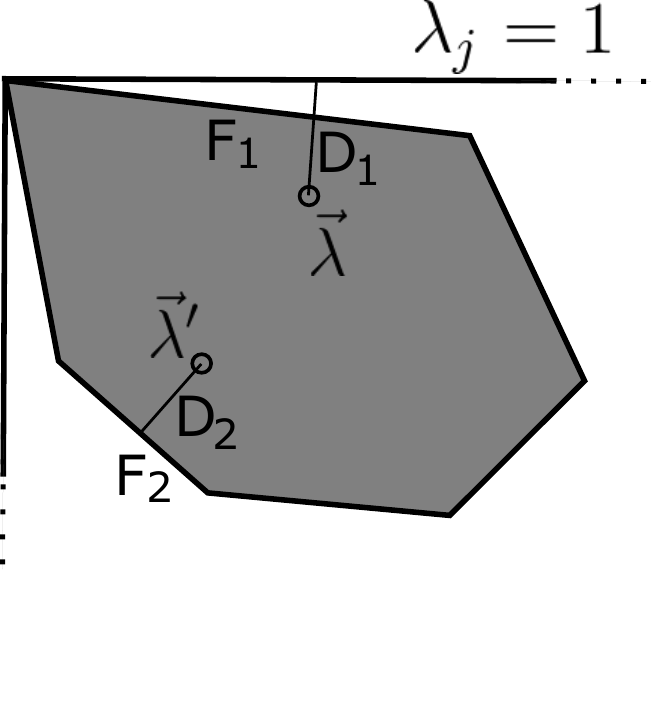}
\caption{Left side: Schematic illustration of the polytope $\mathcal{P}$, which is a proper subset of the light-gray `Pauli simplex' $\Sigma$. The Hartree-Fock point is shown as red dot.
Right side: For some GPC \mbox{($D_1$ but not $D_2$)} a small distance of $\vec{\lambda}$ to the boundary of the surrounding Pauli simplex enforces a small distance $D_j(\vec{\lambda})$ of $\vec{\lambda}$ to the polytope facet $F_j$.}
\label{fig:polytope}
\end{figure}
A crucial point is that the intersection of the polytope's boundary  $\partial \mathcal{P}$ and the boundary $\partial \Sigma$ of the Pauli simplex is not empty. In the following we will show that the set of intersection points exhibits a hierarchical structure which in particular describes the redundancy of the incomplete PC given the more restrictive GPC. As a consequence of the inclusion relation $\mathcal{P}\subset \Sigma$, already observed but not further elaborated in \cite{Mazz14}, one finds for $\vec{\lambda}\in \mathcal{P}$ that $\vec{\lambda}\in \partial \Sigma\Rightarrow  \vec{\lambda}\in \partial \mathcal{P}$.
In particular, this applies to the specific part of the boundary of $\Sigma$ corresponding to pinning by some PC. However, since $\mathcal{P}$ is not only defined by GPC but also by ordering constraints $\lambda_i-\lambda_{i+1}\geq 0$, $\vec{\lambda}\in \partial \mathcal{P}$ does not always correspond to pinning by GPC.
To explore this in more detail, we express the sets of PC in close analogy to Eq~(\ref{eq:gpc}) in the compact form
\begin{equation}\label{eq:PEPs}
S_{r,s}(\vec{\lambda})\equiv\sum_{i=1}^r(1-\lambda_i)+\sum_{j=d+1-s}^d \lambda_j \,\geq\, 0
\end{equation}
with $r\leq N$, $s\leq d-N$, $r+s>0$,
and introduce the corresponding simplicial facets of $\Sigma$
\begin{eqnarray}\label{eq:Sfacets}
\Sigma_{r,s} &\equiv& \{\vec{\lambda}\in \Sigma\,|\,S_{r,s}(\vec{\lambda})=0\}\nonumber \\
&=&\Sigma|_{\tiny
\begin{array}{l}
\lambda_1 = \ldots =\lambda_r=1\\
\lambda_{d+1-s}=\ldots=\lambda_{d}=0
\end{array}}\,.
\end{eqnarray}
Clearly, there is an inclusion relation for those facets:
$\Sigma_{N,d-N}$ is a $0$-dimensional simplex, coinciding with the Hartree-Fock point $\vec{\lambda}_{HF}\equiv(1,\ldots,1,0,\ldots)$. $\Sigma_{N,d-N}$ is contained in the next larger facet, $\Sigma_{N-1,d-N-2}$ which is a $1$-dimensional simplex \footnote{Due to the normalization $\lambda_1+\ldots+\lambda_d=N$ the simplices $\Sigma_{N-1,d-N}$ and $\Sigma_{N,d-N-1}$ coincide with $\Sigma_{N,d-N}$ and we omit them.}. This continues up to the largest two facets $\Sigma_{1,0}$ and $\Sigma_{0,1}$ both of dimension $d-2$.
In general we have
\begin{equation}\label{eq:Sfacethier}
\Sigma_{r,s}\subseteq \Sigma_{r',s'}\quad\Leftrightarrow \quad r \geq r' \wedge s \geq s'\,.
\end{equation}
Similar to $\vec{\lambda} \in F_j$, pinning by some PC, i.e.~$\vec{\lambda} \in \Sigma_{r,s}$, implies strong structural simplifications for the corresponding $N$-fermion state: $r$ fermions are frozen in the first $r$ natural orbitals $\ket{1},\ldots,\ket{r}$ and the last $s$ natural orbitals $\ket{d-s+1},\ldots,\ket{d}$ are inactive, i.e.
\begin{equation}\label{eq:Psi-rs}
\ket{\Psi}= a_1^\dagger\ldots a_r^\dagger\ket{\Psi'}\,.
\end{equation}
The $(N-r)$-fermion state $\ket{\Psi'}$ lives in the corresponding \emph{active space}, $\ket{\Psi'}\in \wedge^{N-r}[\mathcal{H}_1^{(d-r-s)}]$, where $\mathcal{H}_1^{(d-r-s)}$ is spanned by the active orbitals $\ket{r+1},\ldots,\ket{d-s}$.

From a geometrical viewpoint, for any pair $(r,s)$ pinning by the corresponding PC (\ref{eq:PEPs}) implies pinning to the facet $F_j$ of a given GPC $D_j$ if and only if
\begin{equation}\label{PEPtoGPCgeom}
\mathcal{P} \cap \Sigma_{r,s} \subset F_j\,.
\end{equation}
It is worth noticing that $\mathcal{P} \cap \Sigma_{r,s}$ coincides with the polytope $\mathcal{P}'$ for the setting $(N',d')\equiv(N-r,d-r-s)$ (see e.g.~Ref.~\cite{CS2013}). In other words, Eq.~(\ref{PEPtoGPCgeom}) states that $\forall \vec{\lambda}'\in \mathcal{P}'$
\begin{equation}\label{eq:PEPtoGPCD}
D_j(\underbrace{1,\ldots,1}_r,\vec{\lambda}',\underbrace{0,\ldots,0}_{s}) =0\,.
\end{equation}

Based on these observations, natural classes $\mathcal{C}_{r,s}$ of GPC arise,
\begin{equation}\label{eq:classes}
\mathcal{C}_{r,s} \equiv \{D_j\,|\,\mathcal{P} \cap \Sigma_{r,s} \subset F_j\}\,.
\end{equation}
The class $\mathcal{C}_{r,s}$ contains exactly those GPC that are saturated whenever the corresponding PC (\ref{eq:PEPs}) are saturated. Consequently, the classes $\mathcal{C}_{r,s}$ describe the redundancy of the incomplete PC given the more restrictive GPC.
Furthermore, the hierarchy (\ref{eq:Sfacethier}) implies a hierarchy for these classes,
\begin{equation}\label{eq:classeshier}
\mathcal{C}_{r,s}\subseteq \mathcal{C}_{r',s'}\quad\Leftrightarrow \quad r \leq r' \wedge s \leq s'\,,
\end{equation}
and in that sense a partial ordering on $\{(r,s)\}$. By the use of a linear program, we determine all classes and calculate for each GPC $D_j$ all its minimal $(r,s)$ with respect to this partial ordering, i.e.~the smallest pairs $(r,s)$  such that $D_j$ still belongs to the corresponding class $\mathcal{C}_{r,s}$. The results are listed in the `Supplemental Material' for the three cases  $(N,d)=(3,10),(4,10),(5,10)$.
\\

\section{Example for the hierarchy of pinning}\label{sec:example}
\begin{figure}[]
\centering
\includegraphics[width=6.7cm]{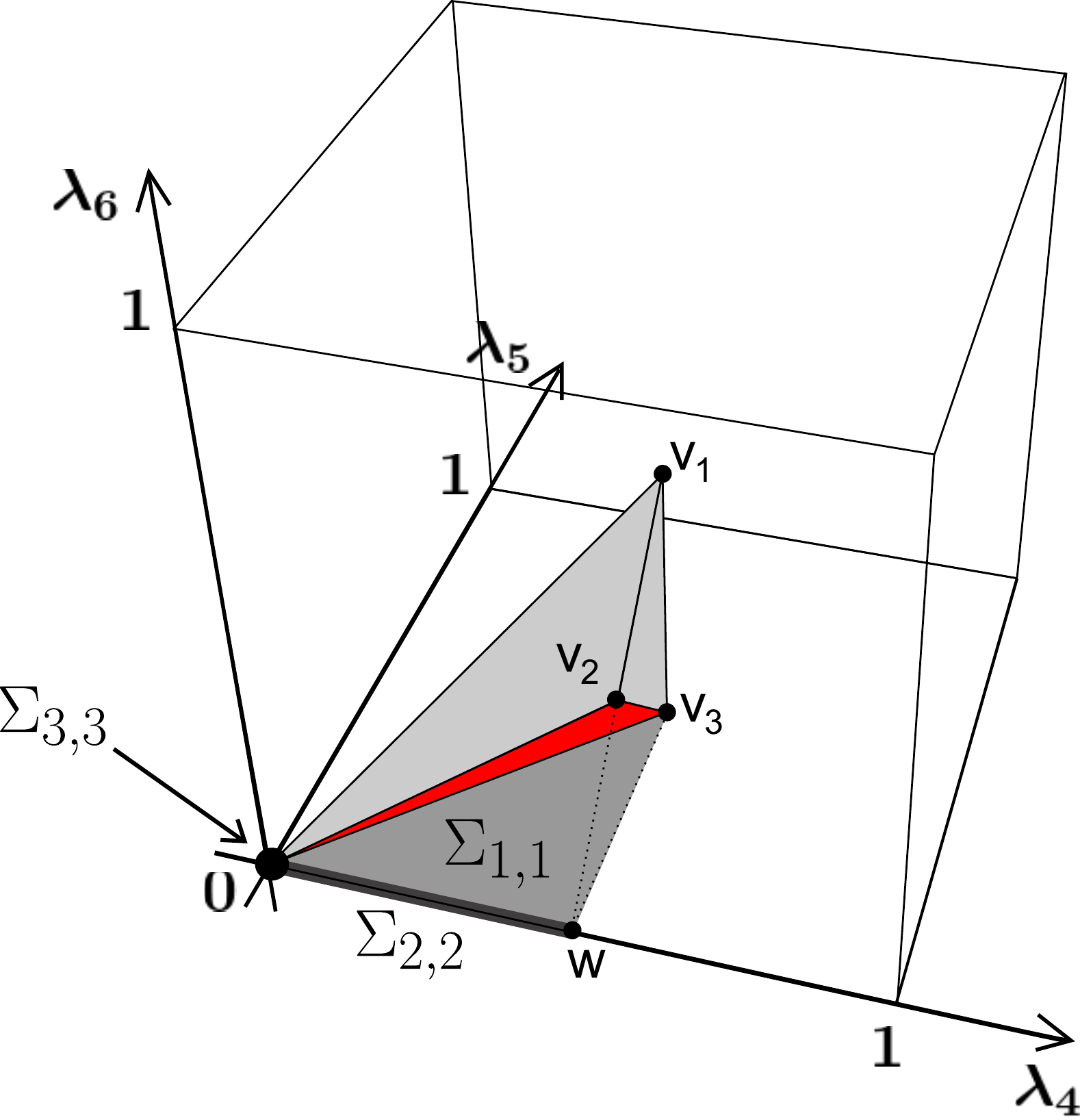}
\caption{For the Borland-Dennis setting, $(N,d)=(3,6)$, the reduced polytope of possible spectra $(\lambda_4,\lambda_5,\lambda_6)$ is shown (the other three NON follow from the equations (\ref{eq:gpc36a}). It is spanned by the vertices $v_{HF}=(0,0,0)$, $v_1=(1/2,1/2,1/2)$, $v_2=(1/2,1/4,1/4)$ and $v_3=(1/2,1/2,0)$. GPC (\ref{eq:gpc36b}) takes the form $\lambda_5+\lambda_6-\lambda_4\geq 0$
and the Pauli Simplex $\Sigma$ is spanned by the vertices $v_{HF}, v_1,v_3$ and $w$.}
\label{fig:polytopeBD}
\end{figure}

We illustrate the hierarchy of simplicial facets of the Pauli simplex $\Sigma$ and the induced
hierarchy of GPC for the particular example of the so-called Borland-Dennis setting, $(N,d)=(3,6)$.
The corresponding polytope is defined by the constraints \cite{Borl1972}
\begin{eqnarray}
&&\lambda_1\geq \lambda_2\geq \lambda_3\geq \lambda_4\geq \lambda_5\geq \lambda_6\geq 0 \label{eq:gpc36ord}\\
&&\lambda_1+\lambda_6=\lambda_2+\lambda_5=\lambda_3+\lambda_4=1\label{eq:gpc36a}\\
&&D(\vec{\lambda})\equiv 2-(\lambda_1+\lambda_2+\lambda_4)\geq 0 \label{eq:gpc36b}\,.
\end{eqnarray}
It is quite special that some GPC take the form of equalities. Due to Eq.~(\ref{eq:gpc36a}) we can choose $\lambda_4,\lambda_5,\lambda_6$ as the independent
variables and present the corresponding reduced polytope of possible vectors $(\lambda_4,\lambda_5,\lambda_6)$ in Fig.~\ref{fig:polytopeBD}.
Also various facets of the Pauli simplex are shown: The smallest one is the $0$-simplex $\Sigma_{3,3}$ given by the Hartree-Fock point $(\lambda_4,\lambda_5,\lambda_6)=(0,0,0)$. It is contained in the next larger facet, the $1$-simplex $\Sigma_{2,2}$ defined by $\lambda_1=\lambda_2=1$, $\lambda_5=\lambda_6=0$. $\Sigma_{2,2}$ is then contained in the largest facet, $\Sigma_{1,1}$, a $2$-simplex shown in gray and defined by $\lambda_6=0$ and $\lambda_5\leq \lambda_4 \leq \frac{1}{2}$. Notice, that it does not make sense to consider the other facets $\Sigma_{r,s}$ with $r\neq s$ since we used the explicit equations (\ref{eq:gpc36a}) to reduce the polytope to three dimensions.
The only GPC which takes the form of a proper inequality is given in this reduced description by $\lambda_5+\lambda_6-\lambda_4\geq 0$. The facet $F_D$ corresponding to pinning is spanned by the Hartree-Fock point and the vertices $v_2,v_3$ and is shown in red. We can now determine whether pinning by this GPC is induced by pinning by some PC.
First, we observe $\Sigma_{3,3}\subset F_D$, i.e.~pinning by all PC implies pinning of GPC (\ref{eq:gpc36b}). Next, we consider the smaller pair $(r,s)=(2,2)$. Clearly, whenever $\vec{\lambda}\in\mathcal{P}$ lies in $\Sigma_{2,2}$ it also lies in the red facet (actually $\vec{\lambda}$ has to coincide with the Hartree-Fock point). Thus, constraint (\ref{eq:gpc36b}) belongs also to the smaller class $\mathcal{C}_{2,2}$. Does it also belong to the smallest class $\mathcal{C}_{1,1}$? Yes, it belongs also to that class since $\vec{\lambda}\in\mathcal{P}\cap\Sigma_{1,1}$ geometrically implies that $\vec{\lambda}$ lies on the line between the Hartree-Fock point and the polytope vertex $v_3$. This line is an edge of the pinning facet $F_D$. Hence, the minimal pair $(r,s)$ for GPC (\ref{eq:gpc36b}) is $(r,s)=(1,1)$.
\\

\section{The measure}\label{sec:measure}
Since pinning by PC implies pinning by specific GPC according to Eq.~(\ref{PEPtoGPCgeom}), the same also holds for quasipinning.
Due to the flat geometry of polytope facets this implication yields \emph{linear} upper bounds on the $l^1$-distance of $\vec{\lambda}$ to the corresponding polytope facets of the form (see also right side of Fig.~\ref{fig:polytope})
\begin{equation}\label{eq:Dvsdist}
\mbox{dist}_1(\vec{\lambda},F_j) \leq \tilde{c}_j \,\mbox{dist}_1(\vec{\lambda},\Sigma_{r,s})\,,
\end{equation}
for all $(r,s)$ with $D_j \in \mathcal{C}_{r,s}$.
Since we are mainly interested in the values $D_j(\vec{\lambda})$ we use
$\mbox{dist}_1(\vec{\lambda},F_j) \propto D_j(\vec{\lambda})$ (see App.~\ref{app:distF} for more details) to obtain bounds of the form
\begin{equation}\label{eq:DvsS}
D_j(\vec{\lambda}) \leq c_j \,\mbox{dist}_1(\vec{\lambda},\Sigma_{r,s})\,.
\end{equation}
Due to the hierarchy (\ref{eq:classeshier}) we consider (\ref{eq:DvsS}) only for the minimal pairs $(r,s)$. The optimal (smallest) prefactors $c_j$
for various GPC and all their minimal $(r,s)$ are determined by a linear program and listed in the `Supplemental Material'.

Equipped with bounds (\ref{eq:DvsS}), we can now decide whether given quasipinning by a GPC follows from possible approximate
saturation of PC or whether it goes beyond that. Whenever $D_j(\vec{\lambda})$ is significantly smaller
than its upper bound (\ref{eq:DvsS}) the quasipining is stronger than one could expect from possible quasipinning by PC. Such nontriviality of quasipinning by GPC which represents the influence of the exchange symmetry \emph{beyond} the exclusion principle is quantified by the ratio of both sides in (\ref{eq:DvsS}). This naturally motivates the following definition for a measure
\begin{equation}\label{eq:Qj}
Q_j(\vec{\lambda}) \equiv -\log_{10}{\big[D_j(\vec{\lambda})/c_j \mbox{dist}_1(\vec{\lambda},\Sigma_{r,s})\big]}\,.
\end{equation}
For GPC with more than one minimal $(r,s)$, i.e.~more than one independent upper bound (\ref{eq:DvsS}), we divide $D_j(\vec{\lambda})$ in Eq.~(\ref{eq:Qj}) by the maximum of all upper bounds. By construction, $Q_j(\vec{\lambda})$ is nonnegative.
For given $Q_j$, the minimal distance of $\vec{\lambda}$ to the polytope boundary is $10^{Q_j}$ times smaller than one could expect from an approximate saturation of PC. Therefore, for $Q_j\lesssim 1$ quasipinning is rather trivial and for $Q_j\gtrsim2$ quite nontrivial.
For practical applications we define the overall $Q$-parameter,
\begin{equation}\label{eq:Q}
Q\equiv \max_j\big(Q_j\big)\,.
\end{equation}

A `Mathematica'-package implementing all $Q$-parameters can be obtained from the authors or may be found in he `Supplemental Material'. The relation between $Q$-parameters of different settings $(N,d)$ is discussed in App.~\ref{app:Qrelation}.
\\

\section{Operational meaning of Q}\label{sec:operational}
In addition to its main purpose of measuring the influence of the fermionic exchange symmetry in the one-particle picture \emph{beyond} Pauli's exclusion principle the $Q$-parameter has also an operational meaning. To elaborate on this in more detail we recall the selection rule (\ref{eq:PinPsi}) of Slater determinants in case of pinning. This remarkable structural simplification of $N$-fermion quantum states holds also (approximately) in case of quasipinning. To be more precise, one finds
\cite{CSQuasipinning,CSHF},
\begin{equation}\label{eq:Dstructure}
\alpha_- D(\vec{\lambda}) \leq 1-\|\hat{P}_{\mathcal{I}_D}\Psi\|_{L^2}^2\leq \alpha_+ \, D(\vec{\lambda})\,,
\end{equation}
for some $\alpha_-,\alpha_+>0$, where in general $\hat{P}_{\mathcal{I}}$ is defined as the operator projecting on the space spanned by $\{\ket{\bd{i}}\}_{\bd{i}\in \mathcal{I}}$.
Estimate (\ref{eq:Dstructure}) also holds for PC $S_{r,s}(\vec{\lambda})\geq 0$, where the corresponding set $\mathcal{I}_{S_{r,s}}$ is given by the configurations $\bd{i}=(1,\ldots,r,i_{r+1},\ldots,i_N)$, $r< i_{\nu} < d-s+1$, containing all $1,\ldots,r$ but none of $d+1-s,\ldots,d$.
Such estimates of the form (\ref{eq:Dstructure}) emphasize the striking relevance that one-particle information can have for the description of many fermion quantum systems. Combining condition (\ref{eq:Dstructure}) for PC $S_{r,s}$ with (\ref{eq:Dstructure}) for a GPC $D_j\in \mathcal{C}_{r,s}$ yields
\begin{equation}\label{eq:DSratio}
\beta_- \cdot10^{-Q_j(\vec{\lambda})} \leq \frac{1-\|\hat{P}_{\mathcal{I}_{D_j}}\Psi\|_{L^2}^{\,2}}{1-\|\hat{P}_{\mathcal{I}_{S_{r,s}}}\Psi\|_{L^2}^{\,2}}\leq \beta_+ \cdot10^{-Q_j(\vec{\lambda})}\,.
\end{equation}

\begin{figure}[]
\centering
\includegraphics[width=8.6cm]{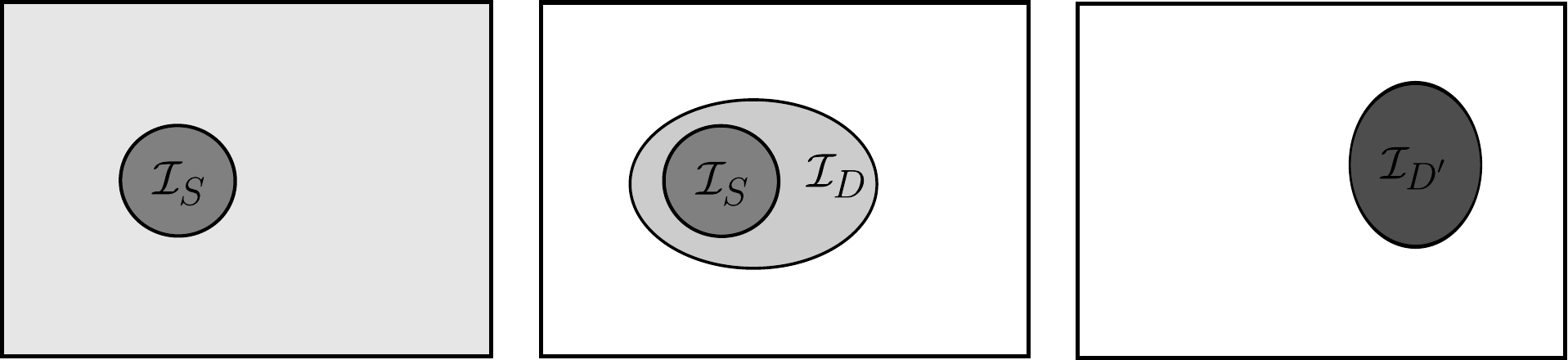}
\caption{Schematic illustration of how the $L^2$-weight of $\ket{\Psi}$ is distributed over various configurations. Higher weight is indicated by darker colors. On the left we use only the information of quasipinning by PC. Stronger implications follow if additional nontrivial quasipinning by GPC is taken into account (middle). Right: even if none of the PC is (approximately) saturated, strong quasipinning by GPC leads to structural simplifications (see also text)}
\label{fig:SR2}
\end{figure}

To explain the significance of estimate (\ref{eq:DSratio}), recall that pinning by GPC and PC is in a one-to-one correspondence to the structure (\ref{eq:PinPsi}) and
(\ref{eq:Psi-rs}), respectively, with corresponding $\mathcal{I}$. Since pinning by PC $S_{r,s}$ implies pinning by GPC $D\in \mathcal{C}_{r,s}$ we have $\mathcal{I}_{S_{r,s}}\subseteq \mathcal{I}_{D_j}$. Therefore, a given $N$-fermion quantum state with NON $\vec{\lambda}$ can be written as
\begin{equation}
\ket{\Psi} \equiv \ket{\Psi_{S}}+ \ket{\Psi_{D\setminus S}}+\ket{\Psi_{R}}
\end{equation}
with $\ket{\Psi_{S}}\equiv \hat{P}_{\mathcal{I}_{S_{r,s}}}\ket{\Psi}$,  $\ket{\Psi_{D\setminus S}}\equiv \hat{P}_{\mathcal{I}_{D_j}\setminus \mathcal{I}_{S_{r,s}}}\ket{\Psi}$ and $\ket{\Psi_{R}}\equiv \hat{P}_{\mathcal{I}_{D_j}^{\,C}}\ket{\Psi}$, where $\mathcal{I}_{D_j}^{\,C}$ denotes the complement of $\mathcal{I}_{D_j}$. Estimate (\ref{eq:DSratio}) then relates the $L^2$-weights of $\ket{\Psi_{D\setminus S}}$ and $\ket{\Psi_{R}}$.
This is illustrated in Fig.~\ref{fig:SR2}. On the left, only quasipinning by PC is considered which implies that most of the $L^2$-weight of $\ket{\Psi}$ is covered by the configurations $\bd{i}\in \mathcal{I}_{S_{r,s}}$. However, given additional \emph{nontrivial} quasipinning by a GPC $D_j\in \mathcal{C}_{r,s}$, $\ket{\Psi}$ can be further specified since almost all weight outside $\mathcal{I}_{S_{r,s}}$ needs to lie in $\mathcal{I}_{D_j}$ (shown in the middle). This means that whenever quasipinning by GPC exceeds quasipinning by PC, Eq.~(\ref{eq:DSratio}) allows one to find much more accurate approximations for $\ket{\Psi}$ than by just freezing $r$ electrons in orbitals $\ket{1},\ldots,\ket{r}$ and omitting $s$ virtual orbitals. This latter case of `large' $Q$ includes the specific case $r=s=0$, i.e.~no approximate saturation of any PC (illustrated on the right side). Note also that depending on the corresponding GPC $D'$, the set $\mathcal{I}_{D'}$ of possible configurations might be even smaller then the set $\mathcal{I}_{S}$ illustrated on the left and in the middle of Fig.~\ref{fig:SR2}.
\\

\section{Proof of principle: Application to a simple model}\label{sec:application}
By applying the $Q$-parameter to a simple model we provide a proof of principle. This will not only emphasize the significance of our developed concepts but also \emph{conclusively} confirm that the fermionic exchange symmetry has a relevance in the one-particle picture \emph{beyond} Pauli's 90 years old exclusion principle. We consider the same simple model as in Ref.~\cite{CS2013},
\begin{equation}\label{eq:Ham}
H_N = \sum_{i=1}^{N} \left(\frac{\vec{p}_i^{\,2}}{2 m} +\frac{m}{2} {\vec{x}_i}^{\,t}\Omega {\vec{x}_i}\right) + \frac{K}{2} \sum_{1\leq i<j\leq N} (\vec{x}_i-\vec{x}_j)^2,
\end{equation}
where $\vec{x}\in \mathbb{R}^n$, $n\in \N$, $\Omega\equiv\mbox{diag}(\omega_1^2,\ldots,\omega_n^2)$ and  $\vec{p}_i \equiv \frac{\hbar}{i}\vec{\nabla}$.

This analytically solvable $N$-Harmonium model has quite a long tradition in physics. Yet, its suggested potential significance for concrete systems as, e.g., for quantum dots \cite{HarmQdots} is here of secondary importance.

\begin{figure}[]
\centering
\includegraphics[width=8.5cm]{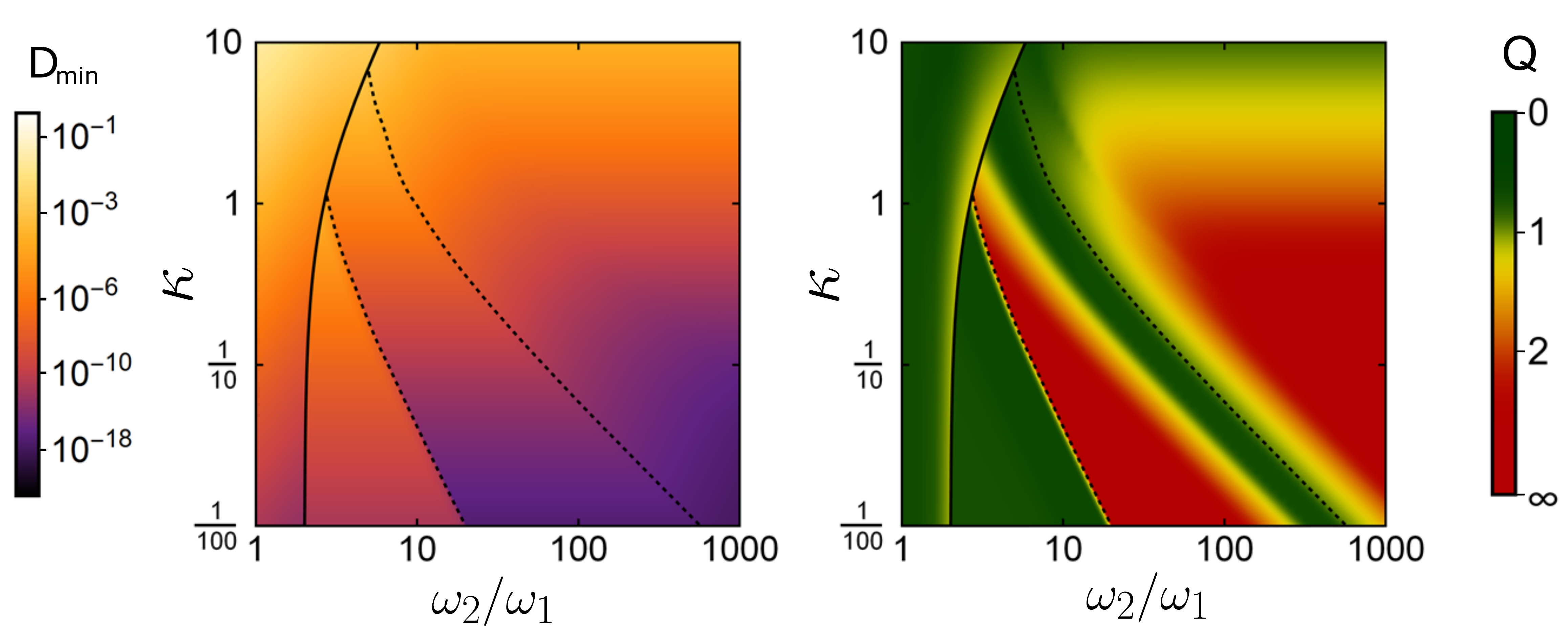}
\caption{Quasipinning results for $3$-Harmonium ground state in two dimensions with external trap frequencies $\omega_1,\omega_2$ and coupling strength $\kappa$. Minimal distance $D_{min}$ of $\vec{\lambda}$ to polytope boundary is shown on the left and its nontriviality is quantified on the right (see text for details).}
\label{fig:2to1D}
\end{figure}

As dimensionless coupling strengths for the different dimensions $i=1,\ldots,n$ we introduce $\kappa_i = \frac{N K}{m \omega_i^2}$. Due to a duality of NON, observed in Ref.~\cite{Nagydual1,CS2013,CS2013NO,Nagydual2} and proven in Ref.~\cite{duality}, we restrict to $\kappa_i>0$. We consider the exemplary case of $N=3$ fermions in $n=2$ spatial dimensions.
The corresponding coupling regime can be parameterized by $\kappa \equiv \kappa_1$ and $\omega_2/\omega_1$ and the results for the ground state are shown in Fig.~\ref{fig:2to1D}. There, the solid black line denotes the crossing of the ground state and the first excited state and dashed lines indicate crossing of NON which can change the quasipinning behavior, as well \cite{CS2015Hubbard}.
From the left side we learn that quasipinning by GPC becomes stronger whenever the coupling $\kappa$ decreases and the trap becomes more anisotropic.
However, the $Q$-parameter presented on the right shows that the phenomenon of quasipinning is much more involved than previously appreciated. For instance, the strong quasipinning in the regime  $\kappa\cdot\omega_2/\omega_1\approx 10$ is completely trivial, i.e.~it is an immediate consequence of quasipinning by PC. It is surrounded from both sides by regimes of highly nontrivial quasipinning by GPC. Similar results are found also for larger $N$ and larger dimensions \cite{CS2016a,CS2016b}. This confirms conclusively the physical relevance of the GPC (recall also Sec.~\ref{sec:pinning}) and shows that the fermionic exchange symmetry can have an influence on the one-particle picture \emph{beyond} Pauli's exclusion principle for concrete systems.

\section{Summary and conclusion}\label{sec:concl}
We have explored whether the fermionic exchange symmetry --- a property on the $N$-particle level --- can have an influence on the one-particle picture \emph{beyond} Pauli's exclusion principle. By exploiting a geometric picture, we first explored and described in the form of a pinning-hierarchy the redundancy of Pauli's incomplete exclusion principle given the more restrictive \emph{generalized Pauli constraints} (GPC). Based on this hierarchy, we succeeded in constructing a measure (`$Q$-parameter') which allows one to quantify for concrete systems the influence of the GPC  \emph{beyond} the Pauli exclusion principle constraints (PC).

From an operational viewpoint, the $Q$-parameter has an even more important additional significance, since it quantifies possible structural simplifications for $N$-fermion quantum states beyond the active space size. In particular, this provides a new perspective on fermionic multipartite correlation since it allows the distinction between static and dynamic correlation. To explain this we first need to discuss the meaning of the PC $S_{r,s}(\vec{\lambda})\geq 0$ (Eq.~(\ref{eq:PEPs})). The value $S_{r,s}(\vec{\lambda})$ represents the weight of the corresponding $N$-fermion quantum state $\ket{\Psi}$ \emph{outside} the active space obtained by freezing $r$ electrons (in the first $r$ natural orbitals) and skipping $s$ virtual states (the last $s$ natural orbitals). As a consequence, $S_{r,s}(\vec{\lambda})$ quantifies \cite{CSHF} the numerical accuracy of the `Complete Active Space Self-Consistent Field'-method (see, e.g., the textbook \cite{QChemJensen}) using the corresponding active space. Since so-called `dynamical correlations' (see, e.g., Ref.~{\cite{ReiherCor}) are attributed to a large number of Slater determinants contributing with small amplitudes to $\ket{\Psi}$, the values $\{S_{r,s}(\vec{\lambda})\}$ measure dynamical correlations. Indeed, if $S_{r,s}(\vec{\lambda})\approx 0$ for some $r,s$ sufficiently close to $N$ and $d-N$, respectively, almost all weight of $\ket{\Psi}$ has to lie in a \emph{small} active space. In contrast to the values $\{S_{r,s}(\vec{\lambda})\}$, the parameters $Q_j$ describe how the weight of the $N$-fermion quantum state distributes over configurations within the different active spaces. This is illustrated in Fig.~\ref{fig:SR2}. On the left side (corresponding to a small $Q_j$) all configurations may in principle contribute to $\ket{\Psi}$, but with small amplitudes (dynamic correlations). This is quite different to the middle, where `large' $Q_j$ implies that the weight shrinks to only a few configurations, reflecting the presence of `static correlations' {\cite{ReiherCor}. Since the type of correlations present in a given system determines whether a \emph{multi-reference} or \emph{single-reference} approach is required, the $Q_j$-parameters may prove useful as an additional tool for this important decision.

By applying the $Q$-parameter to a simple model we provided a proof of principle for the relevance of our new concepts.
Our results confirm \emph{conclusively} the physical relevance of the GPC (recall also Sec.~\ref{sec:pinning}) and show that the fermionic exchange symmetry can have an influence on the one-particle picture \emph{beyond} Pauli's exclusion principle for concrete systems. Our findings also suggest that the phenomenon of quasipinning by GPC is much more involved than previously appreciated. Equipped with our measure $Q$, the ongoing debate on the relevance of GPC may change significantly after reconsidering and reanalyzing \cite{Kly1,CS2013,BenavLiQuasi,Mazz14,CSthesis,MazzOpen,CSQuasipinning,BenavQuasi2,RDMFT,Alex,CS2015Hubbard,RDMFT2,Mazz16}.
Indeed, most of those papers have not contrasted their respective findings with the less restrictive Pauli exclusion principle constraints and have therefore confirmed the physical relevance of the generalized Pauli constraints for the wrong reasons.

Last but not least, our work can be regarded as the starting point of a new research avenue based on Klyachko's breakthrough result solving the one-body pure $N$-representability problem: The systematic analysis and quantification of the influence of the fermionic exchange symmetry \emph{beyond} the well-established influence of Pauli's exclusion principle, possibly also in the \emph{more}-particle picture.

\begin{acknowledgements}
We thank C.\hspace{0.5mm}Benavides-Riveros, M.\hspace{0.5mm}Christandl, D.\hspace{0.5mm}Ebler, D.\hspace{0.5mm}Gross, D.\hspace{0.5mm}Jaksch and B.\hspace{0.5mm}Yadin for helpful discussions. We gratefully acknowledge financial support from
the Friedrich-Naumann-Stiftung and Christ Church Oxford (FT), the Oxford Martin School, the NRF (Singapore), the MoE (Singapore) and the EU Collaborative Project TherMiQ (Grant Agreement 618074)
(VV), the Swiss National Science Foundation (Grant P2EZP2 152190), the Oxford Martin Programme on Bio-Inspired Quantum Technologies and the UK Engineering  and Physical Sciences Research Council (Grant EP/P007155/1) (CS).
\end{acknowledgements}

\appendix

\section{Distance of $\vec{\lambda}$ to polytope facets}\label{app:distF}
Let us first consider the restriction of the Euclidean space $\R^d$ to the hyperplane $V$,
\begin{equation}
V\equiv \{\vec{x}\in \R^d\,|\,x_1+\ldots+x_d=N\}\,,
\end{equation}
where $N$ is the particle number of the setting of interest.
Below, we will restrict these considerations to proper spectra $\vec{x}=(x_i)_{i=1}$, i.e.~to nonnegative and decreasingly-ordered entries $x_i$. Consider now a hyperplane in $V$ defined by
\begin{equation}\label{eq:hyperplane}
E_D\equiv \{\vec{x}\in V\,|\, D(\vec{x})\equiv \kappa^{(0)}+\kappa^{(1)}x_1+\ldots +\kappa^{(d)} x_d = 0\}\,.
\end{equation}
We are interested in deriving an explicit expression for the $l^1$-distance of $\vec{y}\in V$ to the hyperplane $E_D\subset V$. For this, we express the distance of $\vec{y}$ to a set $A\subset V$ as
\begin{equation}\label{eq:l1dist}
\mbox{dist}_1(\vec{y},A) = \sup \big( \big\{r\,|\,\mathcal{B}_1^{(r)}(\vec{y})\cap A = \emptyset\big\} \big)\,,
\end{equation}
where
\begin{equation}\label{eq:l1Ball}
\mathcal{B}_1^{(r)}(\vec{y})\equiv \{\vec{x}\in V\,|\,\|\vec{x}-\vec{y}\|_1\leq r\}
\end{equation}
is the $l^1$-Ball with radius $r>0$ around $\vec{y}$ and $\|\vec{x}\|_1\equiv \sum_{i=1}^d|x_1|$ denotes the $l^1$-norm. Obviously, $\mathcal{B}_1^{(r)}(\vec{y})$
has the shape of a cube intersected with the hyperplane $x_1+\ldots+x_d=N$, i.e.~it is in particular a polytope and has therefore finitely many extremal points, called vertices.

For $\vec{y}\not\in A$ it is instructive to think of (\ref{eq:l1dist}) as the radius $r_0$ that one obtains by further increasing the originally very small radius of $\mathcal{B}_1^{(r)}(\vec{y})$ until one of its points reaches $A$. For  $A=E_D$ we observe that in such a process $E_D$ is first reached by a vertex of $\mathcal{B}_1^{(r)}(\vec{y})$. This follows from the fact that $E_D$ is flat. As an elementary exercise one can determine all the vertices of $\mathcal{B}_1^{(r)}(\vec{y})$, following as
\begin{equation}
\vec{v}_{i,j,\pm}^{(r)}(\vec{y})= \vec{y}\pm \frac{r}{2}(\vec{e}_i-\vec{e}_j)\,,
\end{equation}
where $i,j=1,2,\ldots,d$ with $i< j$ and $\vec{e}_j$ is a unit vector in $j$-direction. Hence, to determine $r_0\equiv \mbox{dist}_1(\vec{y},E_D)$ we need to find roots $r\geq0$ of
\begin{equation}\label{eq:D0vertices}
D(\vec{v}_{i,j,\sigma}^{(r)}(\vec{y}))=D(\vec{y})+\sigma \frac{r}{2} (\kappa^{(i)}-\kappa^{(j)})=0\,
\end{equation}
for various $i,j=1,2,\ldots,d$ with $i< j$ and $\sigma=\pm$. $r_0$ will then be given by the minimum of all those solutions.
For a consistency check, notice that $r=0$ solves (\ref{eq:D0vertices}) for $\vec{y}\in E_D$ independent of $i,j,\sigma$ and thus $\mbox{dist}_1(\vec{y},E_D)=0$. If $\vec{y}\not\in E_D$ we have $D(\vec{y})\neq 0$. For $i,j$ with $\kappa^{(i)}=\kappa^{(j)}$ there is no solution of (\ref{eq:D0vertices}). For $i,j$ with $\kappa^{(i)}\neq\kappa^{(j)}$ we find a solution $r>0$ whenever $\sigma$ has the correct sign:
\begin{equation}\label{eq:D0roots}
r= \left|\frac{2 D(\vec{y})}{\kappa^{(i)}-\kappa^{(j)}}\right| \,.
\end{equation}
Finally, this leads to
\begin{equation}\label{eq:distEDfinal}
\mbox{dist}_1(\vec{y},E_D)\equiv r_0=\frac{2D(\vec{y})}{\Delta\kappa_{max}}\quad,\,\forall \vec{y}\in V\,,
\end{equation}
where
\begin{equation}
\Delta\kappa_{max}=\max_{i,j} \big\{|\kappa_i-\kappa_j| \big\}\,.
\end{equation}
In particular, the result (\ref{eq:distEDfinal}) holds for $D$ a GPC and for all $\vec{y}\in V$ and therefore in particular also for all $\nolinebreak{\vec{y} \in \mathcal{P} \subset V}$. By restricting the hyperplane $E_D$ to the polytope $\mathcal{P}$ of possible spectra $\vec{\lambda}\equiv(\lambda_i)_{i=1}^d$ (i.e.~in particular to decreasingly ordered and nonnegative $\lambda_i$) we obtain a facet $F_D$ of that polytope. The $l^1$-distance of $\vec{\lambda} \in \mathcal{P}$ to $F_D$ is identical to the expression (\ref{eq:distEDfinal}) only for those $\vec{\lambda}$ whose minimal distance to $E_D$ is attained in the subset $F_D\subset E_D$. However, this is always the case for $\vec{\lambda}\in \mathcal{P}$ sufficiently close to $F_D$ and $E_D$, respectively. To summarize, whenever $\vec{\lambda}\in \mathcal{P}$ is sufficiently strongly quasipinned to a facet $F_D$ its $l^1$-distance is related to $D(\vec{\lambda})$ by
\begin{equation}\label{eq:distFDfinal}
\mbox{dist}_1(\vec{\lambda},F_D)= \frac{2D(\vec{\lambda})}{\Delta \kappa_{max}}\,.
\end{equation}
For the case that this distance is not small enough, i.e.~in particular larger than the distance of $\vec{\lambda}$ to another facet $F'$ of $\mathcal{P}$ adjacent to $F_D$
the right hand side of (\ref{eq:distFDfinal}) needs to be modified by an additional prefactor larger than one. However, it is important to keep in mind that the value $D(\vec{\lambda})$ is the relevant quantity. Still, for didactic reasons it is instructive and convenient to talk about the polytope $\mathcal{P}$ formed by all possible spectra $\vec{\lambda}$ and relate $D(\vec{\lambda})$ to the distance of $\vec{\lambda}$ to the corresponding polytope facet $F_D$ rather than to $E_D$.
\\

\section{Relation between $Q$-parameters of different settings}\label{app:Qrelation}
In this section we briefly recall the required concept of truncation and comment on the relation between $Q$-parameters of different settings. We consider two different settings, $(N,d)$ and $(N',d')\equiv (N-\Delta N,d-\Delta d)$. Their polytopes $\mathcal{P}$ and $\mathcal{P}'$ are strongly related according to Ref.~\cite{CSthesis},
\begin{equation}\label{eq:polytopesrelat}
\mathcal{P}'=\mathcal{P}\cap\Sigma_{\Delta N, \Delta d-\Delta N}\,.
\end{equation}
This means that intersecting the polytope $\mathcal{P}$ of the larger setting with the
hyperplane defined by
\begin{equation}\label{eq:hyperplane2}
\lambda_1=\ldots=\lambda_{\Delta N}=1\,\,\,\wedge\,\,\,\lambda_{d+1-\Delta d+\Delta N}=\ldots=\lambda_d=0
\end{equation}
yields the poyltope $\mathcal{P}'$ (which is of course still embedded in $\R^d$). Relation (\ref{eq:polytopesrelat}) means that for every GPC $D'$ of the smaller setting there exists at least one corresponding GPC $D$ of the larger setting, an extension of $D'$, i.e.
\begin{eqnarray}\label{eq:Dextension}
D(\vec{\lambda}) &=& \sum_{i=1}^{\Delta N} \tilde{\kappa}_i (1-\lambda_i) + D'(\lambda_{\Delta N +1},\ldots,\lambda_{d-\Delta d +\Delta N}) \nonumber \\
 &&+\sum_{j=d+1-\Delta d+\Delta N}^d \kappa_j \lambda_j\,.
\end{eqnarray}
Eq.~(\ref{eq:Dextension}) means nothing else than
\begin{equation}\label{eq:Dprojected}
D'(\vec{\lambda}') = D(\underbrace{1,\ldots,1}_{\Delta N},\vec{\lambda}',\underbrace{0,\ldots,0}_{d-\Delta d+\Delta N})\,.
\end{equation}
For all remaining GPC $D$ of $(N,d)$ without a partner GPC in $(N',d')$ their restriction to the hyperplane (\ref{eq:hyperplane2})
leads to an inequality in the remaining NON $\vec{\lambda}'$ which linearly depends on the GPC of $(N',d')$. As a consequence, those
GPC $D$ can be neglected whenever the first $\Delta N$ NON are sufficiently close to $1$ and the last $(\Delta d-\Delta N)$ NON are sufficiently
close to $0$.

The concept of a truncated quasipinning analysis follows from those observations: The analysis of possible quasipinning
of given $\vec{\lambda}\in \mathcal{P}$ can be simplified by omitting various NON very close to $1$ and $0$ and exploring quasipinning
for the remaining NON $\vec{\lambda}'$ in the truncated polytope $\mathcal{P}'$. The minimal distance of $\vec{\lambda}$ to a GPC-facet
of $\mathcal{P}$ follows then as the minimal distance of $\vec{\lambda}'$ to a GPC-facet of $\mathcal{P}'$ up to a small error. This error is given by a linear form in the neglected NON, $1-\lambda_{1},\ldots,1-\lambda_{\Delta N}$, $\lambda_{d+1-\Delta d+\Delta N},\ldots,\lambda_d$. Notice, that the concept of truncation allows one to perform a quasipinning analysis even if the polytope $\mathcal{P}$ is not known yet, provided that sufficiently many NON are very close to $1$ and $0$, respectively. This then leads to a truncated setting $(N',d')$, whose polytope might already be known.

The concept of a truncated quasipinning analysis was successfully applied to various states of the $N$-Harmonium (\ref{eq:Ham}) studied in this work.
There, the one-particle Hilbert space is even infinite-dimensional. However, since almost all NON are sufficiently close to $0$, we could truncate the analysis to the settings $(3,10)$, whose GPC are already known. The corresponding truncation error was checked to be always smaller than various distances of the truncated $\vec{\lambda}'$ to the GPC-facets of $\mathcal{P}'$.

The concept of truncation also suggests that the $Q$-parameters of different settings are related.
One may expect the relation
\begin{equation}\label{eq:Qrelation}
Q(\underbrace{1^-,\ldots,1^-}_{\Delta N},\vec{\lambda}',\underbrace{0^+,\ldots,0^+}_{\Delta d-\Delta N}) = Q'(\vec{\lambda}')\,\,\,,\forall \vec{\lambda}'\in \mathcal{P}'\,,
\end{equation}
where $1^-, 0^+$ means to perform a corresponding limit (avoiding possible expressions of the form $0/0$). However, since some of the individual prefactors in (\ref{eq:DvsS}) may vary slightly for larger settings, (\ref{eq:Qrelation}) may hold in those cases only approximately.
Yet, this would not have any qualitative influence on the conclusion of trivial or nontrivial quasipinning.
To elaborate on this, we consider two GPC $D'$ in $(N',d')$ and $D$ in $(N,d)$ which are related according to Eq.~(\ref{eq:Dextension}).
Relation (\ref{eq:Dextension}) then implies
\begin{equation}\label{eq:classesrelation}
\forall r,s:\quad D' \in \mathcal{C}_{r,s}' \quad \Leftrightarrow \quad D \in \mathcal{C}_{r+\Delta N,s+\Delta d-\Delta N} \,,
\end{equation}
where $\mathcal{C}_{r,s}'$ are the classes (\ref{eq:classes}) for the setting $(N',d')$ and  $\mathcal{C}_{r,s}$ those for $(N,d)$.
From Eq.~(\ref{eq:classesrelation}) we conclude that a pair $(r',s')$ is minimal (given $D'$ still in $\mathcal{C}_{r',s'}'$)
is equivalent to $(r,s)\equiv (r'+\Delta N,s'+\Delta d -\Delta N)$ minimal (given $D$ still in $\mathcal{C}_{r,s}$). This yields
related optimal bounds (recall (\ref{eq:Dvsdist}) and (\ref{eq:DvsS})),
\begin{eqnarray}
D'(\vec{\lambda}') &\leq& c' \, \mbox{dist}_1(\vec{\lambda}',\Sigma_{r,s}')  \label{eq:DvsSrelation1} \\
D(\vec{\lambda}) &\leq& c \, \mbox{dist}_1(\vec{\lambda},\Sigma_{r+\Delta N,s+\Delta d -\Delta N})\,. \label{eq:DvsSrelation2}
\end{eqnarray}
By restricting the bound for $D(\vec{\lambda})$ in Eq.~(\ref{eq:DvsSrelation2}) to the hyperplane (\ref{eq:hyperplane2})
yields bound~(\ref{eq:DvsSrelation1}) up to a potentially different prefactor. Indeed, we observe
\begin{eqnarray}
c &\equiv& \min_{\vec{\lambda}\in \mathcal{P}} \left(\frac{D(\vec{\lambda})}{\mbox{dist}_1(\vec{\lambda},\Sigma_{r+\Delta N,s+\Delta d -\Delta N})}\right) \nonumber \\
&\leq & \min_{\vec{\lambda}\in \mathcal{P}\cap \Sigma_{\Delta N, \Delta d -\Delta N}} \left(\frac{D(\vec{\lambda})}{\mbox{dist}_1(\vec{\lambda},\Sigma_{r+\Delta N,s+\Delta d -\Delta N})}\right) \nonumber \\
&=&\min_{\vec{\lambda}'\in \mathcal{P}'} \left(\frac{D'(\vec{\lambda}')}{\mbox{dist}_1(\vec{\lambda}',\Sigma_{r,s}')}\right) \nonumber \\
&\equiv& c'\,,
\end{eqnarray}
where we used Eqs.~(\ref{eq:polytopesrelat}), (\ref{eq:Dextension}) and (\ref{eq:Dprojected}) in the second last line.

Whenever the prefactors $c$ and $c'$  are different
the corresponding $Q$-parameters (\ref{eq:Q}), $Q$ for $D$ and $Q'$ for $D'$, do not obey Eq.~(\ref{eq:Qrelation}).
As a consequence, the overall $Q$-parameters (\ref{eq:Q}) may also not fulfill relation (\ref{eq:Qrelation}). Yet, this would not have any qualitative influence on the conclusion of trivial or nontrivial quasipinning. This is based on the fact that $c$ and $c'$ differ only for a few related GPC $D'$ and $D$ and then only by a few percent.

\section{Explicit form of the $Q$-parameter}
The generalized Pauli constraints are known so far only for settings $(N,d)$ with $d\leq10$. We present for the three largest ones, $(N,d)=(3,10),(4,10),(5,10)$, the details of the $Q$-parameter and list them in the `Supplemental Material'. We expressed the generalized Pauli constraints in the canonical form
\begin{equation}\label{eq:gpcapp}
D_j(\vec{\lambda}) \equiv \kappa_j^{(0)}+\sum_{i=1}^d\kappa_j^{(i)} \lambda_i\geq 0\,,
\end{equation}
where $j=1,2,\ldots,\nu_{N,d} < \infty$. For each generalized Pauli constraint $D_j$ we determine by resorting to a linear program their extremal classes $\mathcal{C}_{r,s}$ (10), i.e.~the minimal pairs $(r,s)$ provided $D_j \in\mathcal{C}_{r,s}$ . These pairs $(r,s)$ are presented in the `Supplemental Material', in the second last column of the respective tables. There, we decode these pairs in the form $\{r,d+1-s\}$ and also skip possible entries $r=0$ and $d+1-s=d+1$. Notice, that a single entry $\{t\}$ then stands for $\{0,t\}$ if $t>N$ and for $\{t,d+1\}$ if $t\leq N$.
Some generalized Pauli constraints do not belong to any class $\mathcal{C}_{r,s}$, i.e.~the corresponding polytope facet does not contain the Hartree-Fock point. This is indicate by leaving the corresponding entry in the second last column empty. For some generalized Pauli constraints there is more than one minimal `pair' (but never more than two). Each such minimal `pair' gives rise to an upper bound of the form
\begin{equation}\label{eq:DvsSapp}
D_j(\vec{\lambda}) \leq c_j\,\mbox{dist}_1(\vec{\lambda},\Sigma_{r,s})\qquad\,\forall \vec{\lambda}\in \mathcal{P}\,,
\end{equation}
where $\Sigma_{r,s}$ is defined in Eq.~(8). By using similar techniques as for the derivation of the $l^1$-distance of $\vec{\lambda}$ to the polytope facet $F_D$ (c.f App.~\ref{app:distF}) one shows
\begin{equation}
\mbox{dist}_1(\vec{\lambda},\Sigma_{r,s}) = 2\,\max \Big(\sum_{i=1}^r(1-\lambda_i),\sum_{j=d+1-s}^d\lambda_j\Big)\,.
\end{equation}
The minimal coefficients $c_j$ in Eq.~(\ref{eq:DvsSapp}) are determined by a linear program and are listed in the last columns of the three tables in the `Supplemental Material'. For any generalized Pauli constraint $D_i$ which does not belong to any class $\mathcal{C}_{r,s}$ we establish an elementary upper bound of the form
\begin{equation}
D_i(\vec{\lambda})\leq c_i\cdot 1\,,
\end{equation}
and the corresponding minimal $c_i$ are listed in the tables below as well.

Let us consider an example: The seventh generalized Pauli constraint of the setting $(3,10)$
has two independent upper bounds, described by the two minimal `pairs' $\{r,d+1-s\}=\{7\},\{1,8\}$. The corresponding prefactors follow as $1$ and $3/4$. This means
that the optimal bounds on $D_7$ are given by
\begin{eqnarray}
D_7(\vec{\lambda}) &\leq & 1\cdot\mbox{dist}_1(\vec{\lambda},\Sigma_{0,4})\quad\mbox{and} \nonumber \\
D_7(\vec{\lambda}) & \leq &\frac{3}{4}\cdot\mbox{dist}_1(\vec{\lambda},\Sigma_{1,3}) \,.
\end{eqnarray}
According to the definition (14), the corresponding $Q_7$-parameter follows as,
\begin{equation}
Q_7(\vec{\lambda}) = -\log_{10}{\Big[\frac{D_7(\vec{\lambda})}{\max \big(\mbox{dist}_1(\vec{\lambda},\Sigma_{0,4}),\frac{3}{4}\mbox{dist}_1(\vec{\lambda},\Sigma_{1,3})\big)}\Big]}
\end{equation}
In the same way, we find for the third generalized Pauli constraint of the same setting
\begin{equation}
Q_3(\vec{\lambda}) =-\log_{10}{\Big[\frac{14}{9}D_3(\vec{\lambda})\Big]}\,.
\end{equation}

\bibliography{bibliography}

\begin{thebibliography}{37}%
\makeatletter
\providecommand \@ifxundefined [1]{%
 \@ifx{#1\undefined}
}%
\providecommand \@ifnum [1]{%
 \ifnum #1\expandafter \@firstoftwo
 \else \expandafter \@secondoftwo
 \fi
}%
\providecommand \@ifx [1]{%
 \ifx #1\expandafter \@firstoftwo
 \else \expandafter \@secondoftwo
 \fi
}%
\providecommand \natexlab [1]{#1}%
\providecommand \enquote  [1]{``#1''}%
\providecommand \bibnamefont  [1]{#1}%
\providecommand \bibfnamefont [1]{#1}%
\providecommand \citenamefont [1]{#1}%
\providecommand \href@noop [0]{\@secondoftwo}%
\providecommand \href [0]{\begingroup \@sanitize@url \@href}%
\providecommand \@href[1]{\@@startlink{#1}\@@href}%
\providecommand \@@href[1]{\endgroup#1\@@endlink}%
\providecommand \@sanitize@url [0]{\catcode `\\12\catcode `\$12\catcode
  `\&12\catcode `\#12\catcode `\^12\catcode `\_12\catcode `\%12\relax}%
\providecommand \@@startlink[1]{}%
\providecommand \@@endlink[0]{}%
\providecommand \url  [0]{\begingroup\@sanitize@url \@url }%
\providecommand \@url [1]{\endgroup\@href {#1}{\urlprefix }}%
\providecommand \urlprefix  [0]{URL }%
\providecommand \Eprint [0]{\href }%
\providecommand \doibase [0]{http://dx.doi.org/}%
\providecommand \selectlanguage [0]{\@gobble}%
\providecommand \bibinfo  [0]{\@secondoftwo}%
\providecommand \bibfield  [0]{\@secondoftwo}%
\providecommand \translation [1]{[#1]}%
\providecommand \BibitemOpen [0]{}%
\providecommand \bibitemStop [0]{}%
\providecommand \bibitemNoStop [0]{.\EOS\space}%
\providecommand \EOS [0]{\spacefactor3000\relax}%
\providecommand \BibitemShut  [1]{\csname bibitem#1\endcsname}%
\let\auto@bib@innerbib\@empty
\bibitem [{\citenamefont {Pauli}(1925)}]{Pauli1925}%
  \BibitemOpen
  \bibfield  {author} {\bibinfo {author} {\bibfnamefont {W.}~\bibnamefont
  {Pauli}},\ }\bibfield  {title} {\enquote {\bibinfo {title} {{\"U}ber den
  {Z}usammenhang des {A}bschlusses der {E}lektronengruppen im {A}tom mit der
  {K}omplexstruktur der {S}pektren},}\ }\href
  {http://dx.doi.org/10.1007/BF02980631} {\bibfield  {journal} {\bibinfo
  {journal} {Z. Phys.}\ }\textbf {\bibinfo {volume} {31}},\ \bibinfo {pages}
  {765--783} (\bibinfo {year} {1925})}\BibitemShut {NoStop}%
\bibitem [{\citenamefont {Dyson}(1967)}]{Dyson1967}%
  \BibitemOpen
  \bibfield  {author} {\bibinfo {author} {\bibfnamefont {F.J.}\ \bibnamefont
  {Dyson}},\ }\bibfield  {title} {\enquote {\bibinfo {title} {Ground‐state
  energy of a finite system of charged particles},}\ }\href {\doibase
  http://dx.doi.org/10.1063/1.1705389} {\bibfield  {journal} {\bibinfo
  {journal} {J. Math. Phys.}\ }\textbf {\bibinfo {volume} {8}},\ \bibinfo
  {pages} {1538--1545} (\bibinfo {year} {1967})}\BibitemShut {NoStop}%
\bibitem [{\citenamefont {Lieb}(1976)}]{LiebStab}%
  \BibitemOpen
  \bibfield  {author} {\bibinfo {author} {\bibfnamefont {E.H.}\ \bibnamefont
  {Lieb}},\ }\bibfield  {title} {\enquote {\bibinfo {title} {The stability of
  matter},}\ }\href@noop {} {\bibfield  {journal} {\bibinfo  {journal} {Rev.
  Mod. Phys.}\ }\textbf {\bibinfo {volume} {48}},\ \bibinfo {pages} {553--569}
  (\bibinfo {year} {1976})}\BibitemShut {NoStop}%
\bibitem [{\citenamefont {Dirac}(1926)}]{Dirac1926}%
  \BibitemOpen
  \bibfield  {author} {\bibinfo {author} {\bibfnamefont {P.~A.~M.}\
  \bibnamefont {Dirac}},\ }\bibfield  {title} {\enquote {\bibinfo {title} {On
  the theory of quantum mechanics},}\ }\href {\doibase 10.1098/rspa.1926.0133}
  {\bibfield  {journal} {\bibinfo  {journal} {Proc. R. Soc. A}\ }\textbf
  {\bibinfo {volume} {112}},\ \bibinfo {pages} {661--677} (\bibinfo {year}
  {1926})}\BibitemShut {NoStop}%
\bibitem [{\citenamefont {Heisenberg}(1926)}]{Heis1926}%
  \BibitemOpen
  \bibfield  {author} {\bibinfo {author} {\bibfnamefont {W.}~\bibnamefont
  {Heisenberg}},\ }\bibfield  {title} {\enquote {\bibinfo {title}
  {Mehrk\"orperproblem und {R}esonanz in der {Q}uantenmechanik},}\ }\href
  {\doibase 10.1007/BF01397160} {\bibfield  {journal} {\bibinfo  {journal} {Z.
  Phys.}\ }\textbf {\bibinfo {volume} {38}},\ \bibinfo {pages} {411--426}
  (\bibinfo {year} {1926})}\BibitemShut {NoStop}%
\bibitem [{\citenamefont {Laughlin}(1983)}]{Laughlin1983}%
  \BibitemOpen
  \bibfield  {author} {\bibinfo {author} {\bibfnamefont {R.~B.}\ \bibnamefont
  {Laughlin}},\ }\bibfield  {title} {\enquote {\bibinfo {title} {Anomalous
  quantum {H}all effect: An incompressible quantum fluid with fractionally
  charged excitations},}\ }\href@noop {} {\bibfield  {journal} {\bibinfo
  {journal} {Phys. Rev. Lett.}\ }\textbf {\bibinfo {volume} {50}},\ \bibinfo
  {pages} {1395--1398} (\bibinfo {year} {1983})}\BibitemShut {NoStop}%
\bibitem [{\citenamefont {Borland}\ and\ \citenamefont
  {Dennis}(1972)}]{Borl1972}%
  \BibitemOpen
  \bibfield  {author} {\bibinfo {author} {\bibfnamefont {R.E.}\ \bibnamefont
  {Borland}}\ and\ \bibinfo {author} {\bibfnamefont {K.}~\bibnamefont
  {Dennis}},\ }\bibfield  {title} {\enquote {\bibinfo {title} {The conditions
  on the one-matrix for three-body fermion wavefunctions with one-rank equal to
  six},}\ }\href {http://stacks.iop.org/0022-3700/5/i=1/a=009} {\bibfield
  {journal} {\bibinfo  {journal} {J. Phys. B}\ }\textbf {\bibinfo {volume}
  {5}},\ \bibinfo {pages} {7} (\bibinfo {year} {1972})}\BibitemShut {NoStop}%
\bibitem [{\citenamefont {Altunbulak}\ and\ \citenamefont
  {Klyachko}(2008)}]{Kly3}%
  \BibitemOpen
  \bibfield  {author} {\bibinfo {author} {\bibfnamefont {M.}~\bibnamefont
  {Altunbulak}}\ and\ \bibinfo {author} {\bibfnamefont {A.}~\bibnamefont
  {Klyachko}},\ }\bibfield  {title} {\enquote {\bibinfo {title} {The {P}auli
  principle revisited},}\ }\href {\doibase 10.1007/s00220-008-0552-z}
  {\bibfield  {journal} {\bibinfo  {journal} {Commun. Math. Phys.}\ }\textbf
  {\bibinfo {volume} {282}},\ \bibinfo {pages} {287--322} (\bibinfo {year}
  {2008})}\BibitemShut {NoStop}%
\bibitem [{\citenamefont {Klyachko}(2006)}]{Kly2}%
  \BibitemOpen
  \bibfield  {author} {\bibinfo {author} {\bibfnamefont {A.}~\bibnamefont
  {Klyachko}},\ }\bibfield  {title} {\enquote {\bibinfo {title} {Quantum
  marginal problem and n-representability},}\ }\href
  {http://stacks.iop.org/1742-6596/36/i=1/a=014} {\bibfield  {journal}
  {\bibinfo  {journal} {J. Phys.}\ }\textbf {\bibinfo {volume} {36}},\ \bibinfo
  {pages} {72} (\bibinfo {year} {2006})}\BibitemShut {NoStop}%
\bibitem [{\citenamefont {Altunbulak}(2008)}]{Altun}%
  \BibitemOpen
  \bibfield  {author} {\bibinfo {author} {\bibfnamefont {M.}~\bibnamefont
  {Altunbulak}},\ }\emph {\bibinfo {title} {The {P}auli principle,
  representation theory, and geometry of flag varieties}},\ \href
  {http://www.thesis.bilkent.edu.tr/0003572} {Ph.D. thesis},\ \bibinfo
  {school} {Bilkent University} (\bibinfo {year} {2008})\BibitemShut {NoStop}%
\bibitem [{\citenamefont {Klyachko}(2009)}]{Kly1}%
  \BibitemOpen
  \bibfield  {author} {\bibinfo {author} {\bibfnamefont {A.}~\bibnamefont
  {Klyachko}},\ }\bibfield  {title} {\enquote {\bibinfo {title} {The {P}auli
  exclusion principle and beyond},}\ }\href {http://arxiv.org/abs/0904.2009}
  {\bibfield  {journal} {\bibinfo  {journal} {arXiv:0904.2009}\ } (\bibinfo
  {year} {2009})}\BibitemShut {NoStop}%
\bibitem [{\citenamefont {Klyachko}(2013)}]{Kly5}%
  \BibitemOpen
  \bibfield  {author} {\bibinfo {author} {\bibfnamefont {A.}~\bibnamefont
  {Klyachko}},\ }\bibfield  {title} {\enquote {\bibinfo {title} {The {P}auli
  principle and magnetism},}\ }\href {http://arxiv.org/abs/1311.5999}
  {\bibfield  {journal} {\bibinfo  {journal} {arXiv:1311.5999}\ } (\bibinfo
  {year} {2013})}\BibitemShut {NoStop}%
\bibitem [{\citenamefont {Schilling}()}]{CSQMath12}%
  \BibitemOpen
  \bibfield  {author} {\bibinfo {author} {\bibfnamefont {C.}~\bibnamefont
  {Schilling}},\ }\enquote {\bibinfo {title} {The quantum marginal problem},}\
  in\ \href {\doibase 10.1142/9789814618144_0010} {\emph {\bibinfo {booktitle}
  {Mathematical Results in Quantum Mechanics}}},\ Chap.~\bibinfo {chapter}
  {-1}, pp.\ \bibinfo {pages} {165--176}\BibitemShut {NoStop}%
\bibitem [{\citenamefont {Schilling}(2015{\natexlab{a}})}]{CS2015Hubbard}%
  \BibitemOpen
  \bibfield  {author} {\bibinfo {author} {\bibfnamefont {C.}~\bibnamefont
  {Schilling}},\ }\bibfield  {title} {\enquote {\bibinfo {title} {Hubbard
  model: Pinning of occupation numbers and role of symmetries},}\ }\href
  {\doibase 10.1103/PhysRevB.92.155149} {\bibfield  {journal} {\bibinfo
  {journal} {Phys. Rev. B}\ }\textbf {\bibinfo {volume} {92}},\ \bibinfo
  {pages} {155149} (\bibinfo {year} {2015}{\natexlab{a}})}\BibitemShut
  {NoStop}%
\bibitem [{\citenamefont {Schilling}\ \emph {et~al.}(2013)\citenamefont
  {Schilling}, \citenamefont {Gross},\ and\ \citenamefont
  {Christandl}}]{CS2013}%
  \BibitemOpen
  \bibfield  {author} {\bibinfo {author} {\bibfnamefont {C.}~\bibnamefont
  {Schilling}}, \bibinfo {author} {\bibfnamefont {D.}~\bibnamefont {Gross}}, \
  and\ \bibinfo {author} {\bibfnamefont {M.}~\bibnamefont {Christandl}},\
  }\bibfield  {title} {\enquote {\bibinfo {title} {Pinning of fermionic
  occupation numbers},}\ }\href {\doibase 10.1103/PhysRevLett.110.040404}
  {\bibfield  {journal} {\bibinfo  {journal} {Phys. Rev. Lett.}\ }\textbf
  {\bibinfo {volume} {110}},\ \bibinfo {pages} {040404} (\bibinfo {year}
  {2013})}\BibitemShut {NoStop}%
\bibitem [{\citenamefont {Tennie}\ \emph
  {et~al.}(2016{\natexlab{a}})\citenamefont {Tennie}, \citenamefont {Ebler},
  \citenamefont {Vedral},\ and\ \citenamefont {Schilling}}]{CS2016a}%
  \BibitemOpen
  \bibfield  {author} {\bibinfo {author} {\bibfnamefont {F.}~\bibnamefont
  {Tennie}}, \bibinfo {author} {\bibfnamefont {D.}~\bibnamefont {Ebler}},
  \bibinfo {author} {\bibfnamefont {V.}~\bibnamefont {Vedral}}, \ and\ \bibinfo
  {author} {\bibfnamefont {C.}~\bibnamefont {Schilling}},\ }\bibfield  {title}
  {\enquote {\bibinfo {title} {Pinning of fermionic occupation numbers:
  {G}eneral concepts and one spatial dimension},}\ }\href {\doibase
  10.1103/PhysRevA.93.042126} {\bibfield  {journal} {\bibinfo  {journal} {Phys.
  Rev. A}\ }\textbf {\bibinfo {volume} {93}},\ \bibinfo {pages} {042126}
  (\bibinfo {year} {2016}{\natexlab{a}})}\BibitemShut {NoStop}%
\bibitem [{\citenamefont {Tennie}\ \emph
  {et~al.}(2016{\natexlab{b}})\citenamefont {Tennie}, \citenamefont {Vedral},\
  and\ \citenamefont {Schilling}}]{CS2016b}%
  \BibitemOpen
  \bibfield  {author} {\bibinfo {author} {\bibfnamefont {F.}~\bibnamefont
  {Tennie}}, \bibinfo {author} {\bibfnamefont {V.}~\bibnamefont {Vedral}}, \
  and\ \bibinfo {author} {\bibfnamefont {C.}~\bibnamefont {Schilling}},\
  }\bibfield  {title} {\enquote {\bibinfo {title} {Pinning of fermionic
  occupation numbers: {H}igher spatial dimensions and spin},}\ }\href {\doibase
  10.1103/PhysRevA.94.012120} {\bibfield  {journal} {\bibinfo  {journal} {Phys.
  Rev. A}\ }\textbf {\bibinfo {volume} {94}},\ \bibinfo {pages} {012120}
  (\bibinfo {year} {2016}{\natexlab{b}})}\BibitemShut {NoStop}%
\bibitem [{\citenamefont {Benavides-Riveros}\ \emph {et~al.}(2013)\citenamefont
  {Benavides-Riveros}, \citenamefont {Gracia-Bondia},\ and\ \citenamefont
  {Springborg}}]{BenavLiQuasi}%
  \BibitemOpen
  \bibfield  {author} {\bibinfo {author} {\bibfnamefont {C.}~\bibnamefont
  {Benavides-Riveros}}, \bibinfo {author} {\bibfnamefont {J.}~\bibnamefont
  {Gracia-Bondia}}, \ and\ \bibinfo {author} {\bibfnamefont {M.}~\bibnamefont
  {Springborg}},\ }\bibfield  {title} {\enquote {\bibinfo {title} {Quasipinning
  and entanglement in the lithium isoelectronic series},}\ }\href {\doibase
  10.1103/PhysRevA.88.022508} {\bibfield  {journal} {\bibinfo  {journal} {Phys.
  Rev. A}\ }\textbf {\bibinfo {volume} {88}},\ \bibinfo {pages} {022508}
  (\bibinfo {year} {2013})}\BibitemShut {NoStop}%
\bibitem [{\citenamefont {Chakraborty}\ and\ \citenamefont
  {Mazziotti}(2014)}]{Mazz14}%
  \BibitemOpen
  \bibfield  {author} {\bibinfo {author} {\bibfnamefont {R.}~\bibnamefont
  {Chakraborty}}\ and\ \bibinfo {author} {\bibfnamefont {D.A.}\ \bibnamefont
  {Mazziotti}},\ }\bibfield  {title} {\enquote {\bibinfo {title} {Generalized
  {P}auli conditions on the spectra of one-electron reduced density matrices of
  atoms and molecules},}\ }\href {\doibase 10.1103/PhysRevA.89.042505}
  {\bibfield  {journal} {\bibinfo  {journal} {Phys. Rev. A}\ }\textbf {\bibinfo
  {volume} {89}},\ \bibinfo {pages} {042505} (\bibinfo {year}
  {2014})}\BibitemShut {NoStop}%
\bibitem [{\citenamefont {Theophilou}\ \emph {et~al.}(2015)\citenamefont
  {Theophilou}, \citenamefont {Lathiotakis}, \citenamefont {Marques},\ and\
  \citenamefont {Helbig}}]{RDMFT}%
  \BibitemOpen
  \bibfield  {author} {\bibinfo {author} {\bibfnamefont {I.}~\bibnamefont
  {Theophilou}}, \bibinfo {author} {\bibfnamefont {N.N.}\ \bibnamefont
  {Lathiotakis}}, \bibinfo {author} {\bibfnamefont {M.}~\bibnamefont
  {Marques}}, \ and\ \bibinfo {author} {\bibfnamefont {N.}~\bibnamefont
  {Helbig}},\ }\bibfield  {title} {\enquote {\bibinfo {title} {Generalized
  {P}auli constraints in reduced density matrix functional theory},}\ }\href
  {http://scitation.aip.org/content/aip/journal/jcp/142/15/10.1063/1.4918346}
  {\bibfield  {journal} {\bibinfo  {journal} {J. Chem. Phys.}\ }\textbf
  {\bibinfo {volume} {142}} (\bibinfo {year} {2015})}\BibitemShut {NoStop}%
\bibitem [{\citenamefont {Benavides-Riveros}\ and\ \citenamefont
  {Springborg}(2015)}]{BenavQuasi2}%
  \BibitemOpen
  \bibfield  {author} {\bibinfo {author} {\bibfnamefont {C.~L.}\ \bibnamefont
  {Benavides-Riveros}}\ and\ \bibinfo {author} {\bibfnamefont {M.}~\bibnamefont
  {Springborg}},\ }\bibfield  {title} {\enquote {\bibinfo {title} {Quasipinning
  and selection rules for excitations in atoms and molecules},}\ }\href
  {\doibase 10.1103/PhysRevA.92.012512} {\bibfield  {journal} {\bibinfo
  {journal} {Phys. Rev. A}\ }\textbf {\bibinfo {volume} {92}},\ \bibinfo
  {pages} {012512} (\bibinfo {year} {2015})}\BibitemShut {NoStop}%
\bibitem [{\citenamefont {Chakraborty}\ and\ \citenamefont
  {Mazziotti}(2016)}]{Mazz16}%
  \BibitemOpen
  \bibfield  {author} {\bibinfo {author} {\bibfnamefont {R.}~\bibnamefont
  {Chakraborty}}\ and\ \bibinfo {author} {\bibfnamefont {D.A.}\ \bibnamefont
  {Mazziotti}},\ }\bibfield  {title} {\enquote {\bibinfo {title} {Role of the
  generalized {P}auli constraints in the quantum chemistry of excited
  states},}\ }\href
  {http://onlinelibrary.wiley.com/doi/10.1002/qua.25120/abstract} {\bibfield
  {journal} {\bibinfo  {journal} {Int. J. Quant. Chem.}\ }\textbf {\bibinfo
  {volume} {116}},\ \bibinfo {pages} {784--790} (\bibinfo {year}
  {2016})}\BibitemShut {NoStop}%
\bibitem [{\citenamefont {Mazziotti}(2016)}]{MazzGPC2RDM}%
  \BibitemOpen
  \bibfield  {author} {\bibinfo {author} {\bibfnamefont {D.A.}\ \bibnamefont
  {Mazziotti}},\ }\bibfield  {title} {\enquote {\bibinfo {title}
  {Pure-$n$-representability conditions of two-fermion reduced density
  matrices},}\ }\href {\doibase 10.1103/PhysRevA.94.032516} {\bibfield
  {journal} {\bibinfo  {journal} {Phys. Rev. A}\ }\textbf {\bibinfo {volume}
  {94}},\ \bibinfo {pages} {032516} (\bibinfo {year} {2016})}\BibitemShut
  {NoStop}%
\bibitem [{\citenamefont {Schilling}(2015{\natexlab{b}})}]{CSQuasipinning}%
  \BibitemOpen
  \bibfield  {author} {\bibinfo {author} {\bibfnamefont {C.}~\bibnamefont
  {Schilling}},\ }\bibfield  {title} {\enquote {\bibinfo {title} {Quasipinning
  and its relevance for $n$-fermion quantum states},}\ }\href {\doibase
  10.1103/PhysRevA.91.022105} {\bibfield  {journal} {\bibinfo  {journal} {Phys.
  Rev. A}\ }\textbf {\bibinfo {volume} {91}},\ \bibinfo {pages} {022105}
  (\bibinfo {year} {2015}{\natexlab{b}})}\BibitemShut {NoStop}%
\bibitem [{Note1()}]{Note1}%
  \BibitemOpen
  \bibinfo {note} {Due to the normalization $\lambda _1+\protect \ldots
  +\lambda _d=N$ the simplices $\Sigma _{N-1,d-N}$ and $\Sigma _{N,d-N-1}$
  coincide with $\Sigma _{N,d-N}$ and we omit them.}\BibitemShut {Stop}%
\bibitem [{\citenamefont {Benavides-Riveros}\ \emph {et~al.}()\citenamefont
  {Benavides-Riveros}, \citenamefont {Schilling},\ and\ \citenamefont
  {Vrana}}]{CSHF}%
  \BibitemOpen
  \bibfield  {author} {\bibinfo {author} {\bibfnamefont {C.}~\bibnamefont
  {Benavides-Riveros}}, \bibinfo {author} {\bibfnamefont {C.}~\bibnamefont
  {Schilling}}, \ and\ \bibinfo {author} {\bibfnamefont {P.}~\bibnamefont
  {Vrana}},\ }\href@noop {} {\enquote {\bibinfo {title} {Extended
  {H}artree-{F}ock method based on generalized {P}auli constraints},}\
  }\bibinfo {note} {In preparation}\BibitemShut {NoStop}%
\bibitem [{\citenamefont {Johnson}\ and\ \citenamefont
  {Payne}(1991)}]{HarmQdots}%
  \BibitemOpen
  \bibfield  {author} {\bibinfo {author} {\bibfnamefont {N.~F.}\ \bibnamefont
  {Johnson}}\ and\ \bibinfo {author} {\bibfnamefont {M.~C.}\ \bibnamefont
  {Payne}},\ }\bibfield  {title} {\enquote {\bibinfo {title} {Exactly solvable
  model of interacting particles in a quantum dot},}\ }\href {\doibase
  10.1103/PhysRevLett.67.1157} {\bibfield  {journal} {\bibinfo  {journal}
  {Phys. Rev. Lett.}\ }\textbf {\bibinfo {volume} {67}},\ \bibinfo {pages}
  {1157--1160} (\bibinfo {year} {1991})}\BibitemShut {NoStop}%
\bibitem [{\citenamefont {Pipek}\ and\ \citenamefont {Nagy}(2009)}]{Nagydual1}%
  \BibitemOpen
  \bibfield  {author} {\bibinfo {author} {\bibfnamefont {J.}~\bibnamefont
  {Pipek}}\ and\ \bibinfo {author} {\bibfnamefont {I.}~\bibnamefont {Nagy}},\
  }\bibfield  {title} {\enquote {\bibinfo {title} {Measures of spatial
  entanglement in a two-electron model atom},}\ }\href {\doibase
  10.1103/PhysRevA.79.052501} {\bibfield  {journal} {\bibinfo  {journal} {Phys.
  Rev. A}\ }\textbf {\bibinfo {volume} {79}},\ \bibinfo {pages} {052501}
  (\bibinfo {year} {2009})}\BibitemShut {NoStop}%
\bibitem [{\citenamefont {Schilling}(2013)}]{CS2013NO}%
  \BibitemOpen
  \bibfield  {author} {\bibinfo {author} {\bibfnamefont {C.}~\bibnamefont
  {Schilling}},\ }\bibfield  {title} {\enquote {\bibinfo {title} {Natural
  orbitals and occupation numbers for harmonium: Fermions versus bosons},}\
  }\href {\doibase 10.1103/PhysRevA.88.042105} {\bibfield  {journal} {\bibinfo
  {journal} {Phys. Rev. A}\ }\textbf {\bibinfo {volume} {88}},\ \bibinfo
  {pages} {042105} (\bibinfo {year} {2013})}\BibitemShut {NoStop}%
\bibitem [{\citenamefont {Glasser}\ and\ \citenamefont
  {Nagy}(2013)}]{Nagydual2}%
  \BibitemOpen
  \bibfield  {author} {\bibinfo {author} {\bibfnamefont {M.L.}\ \bibnamefont
  {Glasser}}\ and\ \bibinfo {author} {\bibfnamefont {I.}~\bibnamefont {Nagy}},\
  }\bibfield  {title} {\enquote {\bibinfo {title} {Exact evaluation of entropic
  quantities in a solvable two-particle model},}\ }\href
  {http://www.sciencedirect.com/science/article/pii/S0375960113006695}
  {\bibfield  {journal} {\bibinfo  {journal} {Phys. Lett. A}\ }\textbf
  {\bibinfo {volume} {377}},\ \bibinfo {pages} {2317} (\bibinfo {year}
  {2013})}\BibitemShut {NoStop}%
\bibitem [{\citenamefont {Schilling}\ and\ \citenamefont
  {Schilling}(2014)}]{duality}%
  \BibitemOpen
  \bibfield  {author} {\bibinfo {author} {\bibfnamefont {C.}~\bibnamefont
  {Schilling}}\ and\ \bibinfo {author} {\bibfnamefont {R.}~\bibnamefont
  {Schilling}},\ }\bibfield  {title} {\enquote {\bibinfo {title} {Duality of
  reduced density matrices and their eigenvalues},}\ }\href
  {http://stacks.iop.org/1751-8121/47/i=41/a=415305} {\bibfield  {journal}
  {\bibinfo  {journal} {J. Phys. A}\ }\textbf {\bibinfo {volume} {47}},\
  \bibinfo {pages} {415305} (\bibinfo {year} {2014})}\BibitemShut {NoStop}%
\bibitem [{\citenamefont {Jensen}(2006)}]{QChemJensen}%
  \BibitemOpen
  \bibfield  {author} {\bibinfo {author} {\bibfnamefont {F.}~\bibnamefont
  {Jensen}},\ }\href@noop {} {\emph {\bibinfo {title} {Introduction to
  Computational Chemistry}}}\ (\bibinfo  {publisher} {John Wiley \& Sons},\
  \bibinfo {year} {2006})\BibitemShut {NoStop}%
\bibitem [{\citenamefont {Boguslawski}\ \emph {et~al.}(2012)\citenamefont
  {Boguslawski}, \citenamefont {Tecmer}, \citenamefont {Legeza},\ and\
  \citenamefont {Reiher}}]{ReiherCor}%
  \BibitemOpen
  \bibfield  {author} {\bibinfo {author} {\bibfnamefont {K.}~\bibnamefont
  {Boguslawski}}, \bibinfo {author} {\bibfnamefont {P.}~\bibnamefont {Tecmer}},
  \bibinfo {author} {\bibfnamefont {\"{O}.}\ \bibnamefont {Legeza}}, \ and\
  \bibinfo {author} {\bibfnamefont {M.}~\bibnamefont {Reiher}},\ }\bibfield
  {title} {\enquote {\bibinfo {title} {Entanglement measures for single- and
  multireference correlation effects},}\ }\href {\doibase 10.1021/jz301319v}
  {\bibfield  {journal} {\bibinfo  {journal} {J. Phys. Chem. Lett.}\ }\textbf
  {\bibinfo {volume} {3}},\ \bibinfo {pages} {3129--3135} (\bibinfo {year}
  {2012})}\BibitemShut {NoStop}%
\bibitem [{\citenamefont {Schilling}(2014)}]{CSthesis}%
  \BibitemOpen
  \bibfield  {author} {\bibinfo {author} {\bibfnamefont {C.}~\bibnamefont
  {Schilling}},\ }\emph {\bibinfo {title} {Quantum marginal problem and its
  physical relevance}},\ \href {\doibase 10.3929/ethz-a-010139282} {Ph.D.
  thesis},\ \bibinfo  {school} {ETH-Z\"urich} (\bibinfo {year}
  {2014})\BibitemShut {NoStop}%
\bibitem [{\citenamefont {Chakraborty}\ and\ \citenamefont
  {Mazziotti}(2015)}]{MazzOpen}%
  \BibitemOpen
  \bibfield  {author} {\bibinfo {author} {\bibfnamefont {R.}~\bibnamefont
  {Chakraborty}}\ and\ \bibinfo {author} {\bibfnamefont {D.A.}\ \bibnamefont
  {Mazziotti}},\ }\bibfield  {title} {\enquote {\bibinfo {title} {Sufficient
  condition for the openness of a many-electron quantum system from the
  violation of a generalized {P}auli exclusion principle},}\ }\href {\doibase
  10.1103/PhysRevA.91.010101} {\bibfield  {journal} {\bibinfo  {journal} {Phys.
  Rev. A}\ }\textbf {\bibinfo {volume} {91}},\ \bibinfo {pages} {010101}
  (\bibinfo {year} {2015})}\BibitemShut {NoStop}%
\bibitem [{\citenamefont {Lopes}(2015)}]{Alex}%
  \BibitemOpen
  \bibfield  {author} {\bibinfo {author} {\bibfnamefont {A.}~\bibnamefont
  {Lopes}},\ }\emph {\bibinfo {title} {Pure univariate quantum marginals and
  electronic transport properties of geometrically frustrated systems}},\ \href
  {https://www.freidok.uni-freiburg.de/fedora/objects/freidok:10057/datastreams/FILE1/content}
  {Ph.D. thesis},\ \bibinfo  {school} {University of Freiburg} (\bibinfo {year}
  {2015})\BibitemShut {NoStop}%
\bibitem [{\citenamefont {Wang}\ and\ \citenamefont {Knowles}(2015)}]{RDMFT2}%
  \BibitemOpen
  \bibfield  {author} {\bibinfo {author} {\bibfnamefont {J.}~\bibnamefont
  {Wang}}\ and\ \bibinfo {author} {\bibfnamefont {P.J.}\ \bibnamefont
  {Knowles}},\ }\bibfield  {title} {\enquote {\bibinfo {title} {Nonuniqueness
  of algebraic first-order density-matrix functionals},}\ }\href {\doibase
  10.1103/PhysRevA.92.012520} {\bibfield  {journal} {\bibinfo  {journal} {Phys.
  Rev. A}\ }\textbf {\bibinfo {volume} {92}},\ \bibinfo {pages} {012520}
  (\bibinfo {year} {2015})}\BibitemShut {NoStop}%
\end{thebibliography}%
\onecolumngrid
\newpage
\begin{center}\Large{\textbf{Supplemental Material for: ``Influence of the}\\}
\Large{\textbf{Fermionic Exchange Symmetry beyond Pauli's Exclusion Principle''}}
\end{center}
\setcounter{equation}{0}
\setcounter{figure}{0}
\setcounter{table}{0}
\makeatletter
\renewcommand{\theequation}{S\arabic{equation}}
\renewcommand{\thefigure}{S\arabic{figure}}
\vspace{-0.5cm}

\subsection*{Setting$(N,d)=(3,10)$}
\renewcommand{\arraystretch}{1.14}
\begin{minipage}{18.0cm}
$\hspace{1.8cm}
\begin{array}{|r||r|r|r|r|r|r|r|r|r|r|r||c|c|}
\hline
GPC &\kappa^{(0)}&\kappa^{(1)}&\kappa^{(2)}&\kappa^{(3)}&\kappa^{(4)}&\kappa^{(5)}&\kappa^{(6)}&\kappa^{(7)}&\kappa^{(8)}&\kappa^{(9)}&\kappa^{(10)}
& \{r,d+1-s\} & c \\ \hline
 1 & 1 & -1 & -1 & 1 & -1 & 1 & 1 & 0 & 0 & 0 & 0 & \text{$\{$1, 6$\}$} & 1 \\ \hline
 2 & 1 & -1 & 0 & 0 & 0 & 0 & 0 & 0 & 0 & 0 & -1 & \text{$\{$1, 10$\}$} & \text{1/2} \\ \hline
 3 & 3 & -2 & 0 & 0 & 0 & -1 & -1 & -1 & -1 & 0 & -2 & & \text{9/14} \\ \hline
 4 & 3 & -2 & -1 & 0 & 0 & 0 & 0 & -1 & -1 & -1 & -2 & \text{$\{$3, 4$\}$} & \text{1/2} \\ \hline
 5 & 3 & -3 & -2 & 2 & 1 & 0 & 0 & 0 & 0 & -1 & -2 & \text{$\{$3, 4$\}$} & \text{3/4} \\ \hline
 6 & 3 & -3 & 1 & 0 & 0 & 0 & 0 & -1 & -2 & 2 & -2 & & 1 \\ \hline
 7 & 2 & -1 & 0 & -1 & -1 & 0 & -1 & 0 & 0 & 0 & 0 & \text{$\{$7$\}$, $\{$1, 8$\}$} & \text{1, 3/4} \\ \hline
 8 & 2 & -1 & -1 & 0 & -1 & 0 & 0 & -1 & 0 & 0 & 0 & \text{$\{$1, 8$\}$} & \text{3/4} \\ \hline
 9 & 2 & -1 & -1 & 0 & 0 & -1 & -1 & 0 & 0 & 0 & 0 & \text{$\{$7$\}$, $\{$1, 8$\}$} & \text{1, 3/4} \\ \hline
 10 & 2 & 0 & -1 & -1 & -1 & -1 & 0 & 0 & 0 & 0 & 0 & \text{$\{$7$\}$} & 1 \\ \hline
 11 & 3 & -2 & 0 & -1 & -2 & 1 & -1 & 0 & 0 & 0 & -1 & \text{$\{$1, 8$\}$} & \text{5/4} \\ \hline
 12 & 3 & -2 & 0 & -1 & -1 & 0 & -2 & 1 & 0 & 0 & -1 & \text{$\{$7$\}$, $\{$1, 8$\}$} & \text{2, 5/4} \\ \hline
 13 & 3 & -1 & 0 & -2 & -2 & 0 & -1 & 1 & 0 & 0 & -1 & \text{$\{$7$\}$} & 2 \\ \hline
 14 & 3 & -2 & -1 & 0 & 0 & 0 & -1 & -1 & -2 & 1 & 0 & \text{$\{$3, 4$\}$} & \text{2/5} \\ \hline
 15 & 3 & -2 & 0 & 0 & -1 & -1 & -2 & 1 & -1 & 0 & 0 & & \text{3/5} \\ \hline
 16 & 3 & -2 & 0 & 0 & -1 & -1 & -1 & 0 & -2 & 1 & 0 & & \text{3/5} \\ \hline
 17 & 3 & -2 & -1 & 0 & 0 & -1 & -2 & 1 & 0 & 0 & -1 & \text{$\{$7$\}$, $\{$1, 8$\}$} & \text{2, 5/4} \\ \hline
 18 & 3 & -1 & -2 & 0 & 0 & -2 & -1 & 1 & 0 & 0 & -1 & \text{$\{$7$\}$} & 2 \\ \hline
 19 & 3 & 0 & -2 & -1 & -1 & -2 & 0 & 1 & 0 & 0 & -1 & \text{$\{$7$\}$} & 2 \\ \hline
 20 & 3 & 0 & -1 & -2 & -2 & -1 & 0 & 1 & 0 & 0 & -1 & \text{$\{$7$\}$} & 2 \\ \hline
  21 & 3 & 0 & -2 & -1 & -2 & -1 & 1 & 0 & 0 & 0 & -1 & \text{$\{$1, 6$\}$} & 1 \\ \hline
 22 & 3 & -2 & -1 & 0 & -2 & 1 & 0 & -1 & 0 & 0 & -1 & \text{$\{$1, 8$\}$} & \text{5/4} \\ \hline
 23 & 3 & -1 & -2 & 0 & -2 & 0 & 1 & -1 & 0 & 0 & -1 & \text{$\{$1, 6$\}$} & 1 \\ \hline
 24 & 3 & -2 & -1 & 0 & -2 & 1 & 0 & 0 & -1 & -1 & 0 & \text{$\{$1, 6$\}$} & 1 \\ \hline
 25 & 3 & -1 & -2 & 0 & -2 & 0 & 1 & 0 & -1 & -1 & 0 & \text{$\{$1, 6$\}$} & 1 \\ \hline
 26 & 3 & -1 & -2 & 0 & -2 & 1 & 0 & -1 & 0 & -1 & 0 & \text{$\{$3, 4$\}$} & \text{1/3} \\ \hline
  27 & 3 & -2 & -2 & 1 & -1 & 0 & 0 & -1 & 0 & 0 & -1 & \text{$\{$1, 8$\}$} & \text{5/4} \\ \hline
 28 & 3 & -2 & -2 & 1 & -1 & 0 & 0 & 0 & -1 & -1 & 0 & \text{$\{$1, 6$\}$} & 1 \\ \hline
 29 & 3 & 0 & -2 & -1 & -1 & -2 & 1 & 0 & 0 & -1 & 0 & \text{$\{$1, 6$\}$} & 1 \\ \hline
 30 & 3 & 0 & -1 & -2 & -2 & -1 & 1 & 0 & 0 & -1 & 0 & \text{$\{$1, 6$\}$} & 1 \\ \hline
  31 & 3 & -2 & -2 & 1 & 0 & -1 & -1 & 0 & 0 & 0 & -1 & \text{$\{$1, 8$\}$} & \text{5/4} \\ \hline
 32 & 3 & -2 & -2 & 1 & 0 & 0 & -1 & -1 & -1 & 0 & 0 & \text{$\{$3, 4$\}$} & \text{2/5} \\ \hline
 33 & 2 & -1 & -2 & 1 & 0 & -1 & 1 & 0 & -1 & 1 & 0 & \text{$\{$1, 6$\}$} & 1 \\ \hline
 34 & 2 & -1 & 0 & -1 & -2 & 1 & 1 & 0 & -1 & 1 & 0 & \text{$\{$1, 6$\}$} & 1 \\ \hline
 35 & 2 & 0 & -1 & -1 & -2 & 1 & 0 & 1 & -1 & 1 & 0 & \text{$\{$3, 4$\}$} & \text{1/2} \\ \hline
 36 & 2 & -1 & -2 & 1 & -1 & 1 & 0 & 0 & -1 & 1 & 0 & \text{$\{$3, 4$\}$} & \text{1/3} \\ \hline
 37 & 1 & 0 & -1 & 0 & -1 & 0 & 1 & 0 & 0 & 0 & 0 & \text{$\{$1, 6$\}$} & \text{1/2} \\ \hline
 38 & 1 & 0 & -1 & 0 & 0 & -1 & 0 & 1 & 0 & 0 & 0 & \text{$\{$7$\}$} & 1 \\ \hline
 39 & 1 & -1 & 0 & 0 & -1 & 1 & 0 & 0 & 0 & 0 & 0 & \text{$\{$1, 10$\}$} & \text{1/2} \\ \hline
 40 & 1 & -1 & 0 & 0 & 0 & 0 & -1 & 1 & 0 & 0 & 0 & \text{$\{$7$\}$, $\{$1, 10$\}$} & \text{1, 1/2} \\ \hline
 41 & 1 & 0 & 0 & -1 & -1 & 0 & 0 & 1 & 0 & 0 & 0 & \text{$\{$7$\}$} & 1 \\ \hline
 42 & 1 & -1 & -1 & 1 & 0 & 0 & 0 & 0 & 0 & 0 & 0 & \text{$\{$1, 10$\}$} & \text{1/2} \\ \hline
\end{array}
$
\end{minipage}

\renewcommand{\arraystretch}{1.14}
\begin{minipage}{18.0cm}
$\hspace{1.7cm}
\begin{array}{|r||r|r|r|r|r|r|r|r|r|r|r||c|c|}
\hline
GPC &\kappa^{(0)}&\kappa^{(1)}&\kappa^{(2)}&\kappa^{(3)}&\kappa^{(4)}&\kappa^{(5)}&\kappa^{(6)}&\kappa^{(7)}&\kappa^{(8)}&\kappa^{(9)}&\kappa^{(10)}
& \{r,d+1-s\} & c \\ \hline
 43 & 1 & -1 & 0 & 0 & 0 & 0 & 0 & 0 & -1 & 1 & 0 & \text{$\{$1, 10$\}$} & \text{1/2} \\ \hline
 44 & 3 & -2 & 0 & 0 & 0 & -1 & -1 & -1 & -2 & 1 & -1 & & \text{9/14} \\ \hline
 45 & 3 & -2 & -2 & 1 & 0 & 0 & 0 & -1 & -1 & -1 & -1 & \text{$\{$3, 4$\}$} & \text{1/2} \\ \hline
 46 & 3 & -2 & -1 & 0 & -1 & 0 & -1 & 0 & -1 & 0 & -1 & \text{$\{$1, 10$\}$} & \text{3/4} \\ \hline
 47 & 3 & -1 & -2 & 0 & -1 & -1 & 0 & 0 & -1 & 0 & -1 & \text{$\{$1, 6$\}$} & \text{1/2} \\ \hline
 48 & 3 & -1 & -1 & -1 & -2 & 0 & 0 & 0 & -1 & 0 & -1 & \text{$\{$1, 6$\}$} & \text{1/2} \\ \hline
 49 & 2 & -1 & -2 & 1 & 0 & -1 & 1 & 0 & 0 & 0 & -1 & \text{$\{$1, 6$\}$} & 1 \\ \hline
 50 & 2 & -1 & 0 & -1 & -2 & 1 & 1 & 0 & 0 & 0 & -1 & \text{$\{$1, 6$\}$} & 1 \\ \hline
 51 & 2 & 0 & -1 & -1 & -2 & 1 & 0 & 1 & 0 & 0 & -1 & \text{$\{$3, 4$\}$} & \text{1/2} \\ \hline
 52 & 3 & -2 & -3 & 2 & 0 & 1 & 0 & -1 & -2 & 2 & 1 & \text{$\{$3, 4$\}$} & \text{2/3} \\ \hline
 53 & 3 & 0 & -1 & -2 & -3 & 2 & 1 & 0 & -2 & 2 & 1 & \text{$\{$3, 4$\}$} & \text{3/4} \\ \hline
 54 & 3 & -3 & 1 & -1 & -2 & 2 & -2 & 2 & 1 & 0 & 0 & \text{$\{$1, 8$\}$} & \text{29/12} \\ \hline
 55 & 3 & -2 & 1 & -2 & -3 & 2 & -1 & 2 & 1 & 0 & 0 & \text{$\{$1, 6$\}$} & 2 \\ \hline
  56 & 3 & -3 & -2 & 2 & 1 & 0 & 0 & -1 & -2 & 2 & 1 & \text{$\{$3, 4$\}$} & \text{17/24} \\ \hline
 57 & 3 & -3 & 1 & 0 & 0 & -1 & -2 & 2 & -2 & 2 & 1 & & \text{6/5} \\ \hline
 58 & 3 & -3 & -2 & 2 & 1 & -1 & -2 & 2 & 1 & 0 & 0 & \text{$\{$1, 8$\}$} & \text{29/12} \\ \hline
 59 & 3 & -2 & -3 & 2 & 1 & -2 & -1 & 2 & 1 & 0 & 0 & \text{$\{$1, 6$\}$} & 2 \\ \hline
 60 & 3 & 1 & -3 & -1 & -2 & -2 & 2 & 2 & 1 & 0 & 0 & \text{$\{$1, 6$\}$} & 2 \\ \hline
 61 & 3 & 1 & -2 & -2 & -3 & -1 & 2 & 2 & 1 & 0 & 0 & \text{$\{$1, 6$\}$} & 2 \\ \hline
 62 & 3 & -2 & -3 & 2 & -2 & 2 & 1 & 0 & -1 & 0 & 1 & \text{$\{$3, 4$\}$} & \text{2/3} \\ \hline
 63 & 3 & -2 & -3 & 2 & -2 & 2 & 1 & -1 & 0 & 1 & 0 & \text{$\{$3, 4$\}$} & \text{2/3} \\ \hline
 64 & 3 & -3 & -2 & 2 & -2 & 2 & 1 & -1 & 1 & 0 & 0 & \text{$\{$1, 8$\}$} & \text{29/12} \\ \hline
 65 & 3 & -2 & -2 & 1 & -3 & 2 & 2 & -1 & 1 & 0 & 0 & \text{$\{$1, 6$\}$} & 2 \\ \hline
 66 & 3 & -2 & -3 & 2 & -2 & 1 & 2 & -1 & 1 & 0 & 0 & \text{$\{$1, 6$\}$} & 2 \\ \hline
 67 & 3 & -3 & -2 & 2 & -2 & 2 & 1 & 0 & 0 & -1 & 1 & \text{$\{$1, 6$\}$} & 2 \\ \hline
 68 & 3 & -2 & -2 & 1 & -3 & 2 & 2 & 0 & 0 & -1 & 1 & \text{$\{$1, 6$\}$} & 2 \\ \hline
 69 & 3 & -2 & -3 & 2 & -2 & 1 & 2 & 0 & 0 & -1 & 1 & \text{$\{$1, 6$\}$} & 2 \\ \hline
 70 & 3 & -4 & -3 & 4 & 2 & 1 & 0 & 1 & 0 & -1 & -2 & \text{$\{$3, 4$\}$} & \text{7/6} \\ \hline
 71 & 3 & -4 & 2 & 1 & 0 & 1 & 0 & -1 & -3 & 4 & -2 & & \text{3/2} \\ \hline
 72 & 4 & -3 & 0 & -1 & 0 & -1 & -2 & 1 & 0 & -1 & -2 & \text{$\{$1, 10$\}$} & 2 \\ \hline
 73 & 4 & -3 & 0 & -1 & -2 & 1 & 0 & -1 & 0 & -1 & -2 & \text{$\{$1, 10$\}$} & 2 \\ \hline
 74 & 4 & -3 & -2 & 1 & 0 & -1 & 0 & -1 & 0 & -1 & -2 & \text{$\{$1, 10$\}$} & 2 \\ \hline
 75 & 4 & -1 & -1 & -2 & -3 & 1 & 0 & 0 & 0 & -1 & -2 & \text{$\{$3, 4$\}$} & \text{8/15} \\ \hline
 76 & 4 & -3 & 0 & -1 & 0 & -1 & 0 & -1 & -2 & 1 & -2 & \text{$\{$1, 10$\}$} & 2 \\ \hline
 77 & 4 & -2 & -3 & 1 & -1 & 0 & 0 & -1 & -2 & 1 & 0 & \text{$\{$3, 4$\}$} & \text{1/2} \\ \hline
 78 & 4 & -1 & -1 & -2 & -3 & 1 & 0 & 0 & -2 & 1 & 0 & \text{$\{$3, 4$\}$} & \text{19/36} \\ \hline
 79 & 4 & -3 & 0 & -1 & -2 & 1 & -2 & 1 & 0 & -1 & 0 & \text{$\{$1, 10$\}$} & 2 \\ \hline
 80 & 4 & -3 & 0 & -1 & -2 & 1 & 0 & -1 & -2 & 1 & 0 & \text{$\{$1, 10$\}$} & 2 \\ \hline
 81 & 4 & -2 & 0 & -2 & -3 & 1 & -1 & 1 & 0 & -1 & 0 & \text{$\{$1, 6$\}$} & \text{3/2} \\ \hline
 82 & 4 & -3 & -2 & 1 & 0 & -1 & 0 & -1 & -2 & 1 & 0 & \text{$\{$1, 10$\}$} & 2 \\ \hline
 83 & 4 & -3 & 0 & -1 & 0 & -1 & -2 & 1 & -2 & 1 & 0 & \text{$\{$1, 10$\}$} & 2 \\ \hline
 84 & 4 & -3 & -2 & 1 & 0 & -1 & -2 & 1 & 0 & -1 & 0 & \text{$\{$1, 10$\}$} & 2 \\ \hline
 85 & 4 & -2 & -3 & 1 & 0 & -2 & -1 & 1 & 0 & -1 & 0 & \text{$\{$1, 6$\}$} & \text{3/2} \\ \hline
 86 & 4 & 0 & -3 & -1 & -2 & -2 & 1 & 1 & 0 & -1 & 0 & \text{$\{$1, 6$\}$} & \text{3/2} \\ \hline
 87 & 4 & 0 & -2 & -2 & -3 & -1 & 1 & 1 & 0 & -1 & 0 & \text{$\{$1, 6$\}$} & \text{3/2} \\ \hline
 88 & 4 & -2 & -3 & 1 & -2 & 1 & 0 & -1 & -1 & 0 & 0 & \text{$\{$3, 4$\}$} & \text{1/2} \\ \hline
 89 & 4 & -3 & -2 & 1 & -2 & 1 & 0 & -1 & 0 & -1 & 0 & \text{$\{$1, 10$\}$} & 2 \\ \hline
 90 & 4 & -2 & -2 & 0 & -3 & 1 & 1 & -1 & 0 & -1 & 0 & \text{$\{$1, 6$\}$} & \text{3/2} \\ \hline
 91 & 4 & -2 & -3 & 1 & -2 & 0 & 1 & -1 & 0 & -1 & 0 & \text{$\{$1, 6$\}$} & \text{3/2} \\ \hline
 92 & 3 & -4 & 2 & 1 & 0 & 1 & 0 & -1 & -2 & 3 & -3 & & \text{3/2} \\ \hline
 93 & 3 & -4 & -2 & 3 & 2 & 1 & 0 & 1 & 0 & -1 & -3 & \text{$\{$3, 4$\}$} & \text{7/6} \\ \hline
\end{array}
$
\end{minipage}

\subsection*{Setting$(N,d)=(4,10)$}
\renewcommand{\arraystretch}{1.14}
\begin{minipage}{18.0cm}
$\hspace{1.7cm}
\begin{array}{|r||r|r|r|r|r|r|r|r|r|r|r||c|c|}
\hline
GPC&\kappa^{(0)}&\kappa^{(1)}&\kappa^{(2)}&\kappa^{(3)}&\kappa^{(4)}&\kappa^{(5)}&\kappa^{(6)}&\kappa^{(7)}&\kappa^{(8)}&\kappa^{(9)}&\kappa^{(10)}
& \{r,d+1-s\} & c \\ \hline
 1 & 1 & -1 & 0 & 0 & 0 & 0 & 0 & 0 & 0 & 0 & 0 & \text{$\{$1$\}$} & \text{1/2} \\ \hline
 2 & 3 & -1 & -1 & -1 & 0 & -1 & 0 & 0 & 0 & -1 & 0 & \text{$\{$2, 7$\}$} & \text{1/2} \\ \hline
 3 & 3 & -1 & -1 & -2 & 1 & -2 & 2 & 1 & 2 & 0 & 0 & \text{$\{$4, 5$\}$} & \text{3/4} \\ \hline
 4 & 3 & -1 & -2 & -1 & 1 & -2 & 2 & 2 & 1 & 0 & 0 & \text{$\{$2, 7$\}$} & 2 \\ \hline
 5 & 3 & -2 & -1 & -2 & 2 & -1 & 2 & 1 & 1 & 0 & 0 & \text{$\{$4, 5$\}$} & \text{3/4} \\ \hline
 6 & 3 & -1 & -2 & -2 & 2 & -1 & 1 & 2 & 1 & 0 & 0 & \text{$\{$2, 7$\}$} & 2 \\ \hline
 7 & 4 & -1 & -2 & -2 & 1 & -3 & 3 & 2 & 2 & 0 & 0 & \text{$\{$4, 5$\}$} & 1 \\ \hline
 8 & 4 & -2 & -2 & -3 & 3 & -1 & 2 & 2 & 1 & 0 & 0 & \text{$\{$4, 5$\}$} & 1 \\ \hline
 9 & 5 & -3 & -1 & -2 & 1 & 0 & -1 & 1 & 0 & -1 & 1 & \text{$\{$2, 7$\}$} & \text{7/4} \\ \hline
 10 & 5 & -2 & -1 & -1 & -1 & -3 & 1 & 1 & 1 & 0 & 0 & \text{$\{$2, 7$\}$} & \text{5/4} \\ \hline
 11 & 5 & -3 & -1 & 0 & -1 & -2 & 1 & 1 & 0 & -1 & 1 & \text{$\{$2, 7$\}$} & \text{7/4} \\ \hline
 12 & 5 & -3 & -1 & -2 & 1 & -1 & 1 & 0 & 0 & -1 & 1 & \text{$\{$4, 5$\}$} & \text{5/8} \\ \hline
 13 & 5 & -2 & -3 & -1 & 1 & -1 & 1 & 0 & 0 & -1 & 1 & \text{$\{$2$\}$} & \text{3/2} \\ \hline
 14 & 5 & -2 & -3 & 0 & 0 & -1 & 1 & -1 & 1 & -1 & 1 & \text{$\{$2$\}$} & \text{3/2} \\ \hline
 15 & 5 & -2 & -3 & -1 & 1 & 0 & 0 & -1 & 1 & -1 & 1 & \text{$\{$2$\}$} & \text{3/2} \\ \hline
 16 & 5 & -3 & 0 & -1 & -1 & -2 & 1 & 0 & 1 & -1 & 1 & \text{$\{$4, 5$\}$} & \text{5/8} \\ \hline
 17 & 5 & -2 & -1 & -3 & 1 & -1 & -1 & 1 & 1 & 0 & 0 & \text{$\{$2, 7$\}$} & \text{5/4} \\ \hline
 18 & 5 & -2 & -3 & -1 & 1 & -1 & 1 & -1 & 1 & 0 & 0 & \text{$\{$2$\}$} & \text{3/2} \\ \hline
 19 & 2 & -1 & 0 & -1 & 0 & -1 & 0 & 1 & 0 & 0 & 0 & \text{$\{$7$\}$} & \text{3/4} \\ \hline
 20 & 2 & -1 & 0 & 0 & -1 & -1 & 0 & 0 & 1 & 0 & 0 & \text{$\{$7$\}$, $\{$1, 8$\}$} & \text{3/4, 2} \\ \hline
 21 & 2 & -1 & -1 & 0 & 0 & 0 & 0 & 0 & 0 & -1 & 1 & \text{$\{$2$\}$} & \text{1/2} \\ \hline
 22 & 2 & -1 & -1 & 0 & 0 & 0 & 0 & -1 & 1 & 0 & 0 & \text{$\{$2$\}$, $\{$1, 8$\}$} & \text{1/2, 2} \\ \hline
 23 & 2 & 0 & -1 & -1 & 0 & -1 & 0 & 0 & 1 & 0 & 0 & \text{$\{$7$\}$} & \text{3/4} \\ \hline
 24 & 2 & -1 & 0 & -1 & 0 & 0 & -1 & 0 & 1 & 0 & 0 & \text{$\{$7$\}$, $\{$1, 8$\}$} & \text{3/4, 2} \\ \hline
 25 & 2 & -1 & -1 & 0 & 0 & -1 & 1 & 0 & 0 & 0 & 0 & \text{$\{$2$\}$} & \text{1/2} \\ \hline
 26 & 2 & -1 & -1 & -1 & 1 & 0 & 0 & 0 & 0 & 0 & 0 & \text{$\{$2$\}$} & \text{1/2} \\ \hline
 27 & 4 & -2 & -1 & -1 & 0 & -1 & 0 & 0 & -1 & 0 & 0 & \text{$\{$2, 9$\}$} & 1 \\ \hline
 28 & 4 & -2 & 0 & -1 & -1 & -1 & -1 & 0 & 0 & 0 & 0 & \text{$\{$7$\}$, $\{$1, 8$\}$} & \text{1/2, 2} \\ \hline
 29 & 4 & -2 & -1 & -1 & 0 & 0 & -1 & -1 & 0 & 0 & 0 & \text{$\{$1, 8$\}$, $\{$2, 9$\}$} & \text{2, 1} \\ \hline
  30 & 4 & -2 & -1 & 0 & -1 & -1 & 0 & -1 & 0 & 0 & 0 & \text{$\{$1, 8$\}$, $\{$2, 9$\}$} & \text{2, 1} \\ \hline
 31 & 4 & -1 & -1 & -1 & -1 & -2 & 0 & 0 & 0 & 0 & 0 & \text{$\{$7$\}$} & \text{1/2} \\ \hline
 32 & 4 & -1 & -1 & -2 & 0 & -1 & -1 & 0 & 0 & 0 & 0 & \text{$\{$7$\}$} & \text{1/2} \\ \hline
 33 & 4 & -1 & -2 & -1 & 0 & -1 & 0 & -1 & 0 & 0 & 0 & \text{$\{$2, 9$\}$} & 1 \\ \hline
 34 & 5 & -2 & -1 & -1 & -1 & -2 & 0 & 0 & 0 & -1 & 0 & \text{$\{$2, 7$\}$} & \text{3/4} \\ \hline
 35 & 5 & -2 & -2 & -1 & 0 & -1 & 0 & -1 & 0 & -1 & 0 & \text{$\{$2$\}$} & \text{3/4} \\ \hline
 36 & 5 & -2 & -1 & -2 & 0 & -1 & -1 & 0 & 0 & -1 & 0 & \text{$\{$2, 7$\}$} & \text{3/4} \\ \hline
 37 & 3 & -2 & -1 & -1 & 1 & -1 & 1 & 1 & 0 & 0 & 0 & \text{$\{$2, 7$\}$} & \text{5/4} \\ \hline
 38 & 3 & -1 & -1 & -1 & 0 & -2 & 1 & 1 & 1 & 0 & 0 & \text{$\{$2, 7$\}$} & 1 \\ \hline
 39 & 3 & -1 & -2 & -1 & 1 & -1 & 1 & 0 & 1 & 0 & 0 & \text{$\{$2, 7$\}$} & \text{5/4} \\ \hline
 40 & 3 & -1 & -1 & -2 & 1 & -1 & 0 & 1 & 1 & 0 & 0 & \text{$\{$2, 7$\}$} & 1 \\ \hline
 41 & 8 & -2 & -3 & -4 & 1 & -5 & 4 & 2 & 3 & 0 & -1 & \text{$\{$4, 5$\}$} & \text{3/2} \\ \hline
 42 & 8 & -2 & -3 & -4 & 1 & -5 & 4 & 3 & 2 & -1 & 0 & \text{$\{$4, 5$\}$} & \text{3/2} \\ \hline
 43 & 8 & -3 & -4 & -5 & 4 & -2 & 2 & 3 & 1 & 0 & -1 & \text{$\{$4, 5$\}$} & \text{3/2} \\ \hline
 44 & 8 & -4 & -3 & -5 & 4 & -2 & 3 & 2 & 1 & 0 & -1 & \text{$\{$4, 5$\}$} & \text{3/2} \\ \hline
 45 & 8 & -3 & -4 & -5 & 4 & -2 & 3 & 2 & 1 & -1 & 0 & \text{$\{$4, 5$\}$} & \text{3/2} \\ \hline
 46 & 8 & -2 & -4 & -3 & 1 & -5 & 4 & 3 & 2 & 0 & -1 & \text{$\{$4, 5$\}$} & \text{3/2} \\ \hline
\end{array}
$
\end{minipage}

\renewcommand{\arraystretch}{1.14}
\begin{minipage}{18.0cm}
$\hspace{1.7cm}
\begin{array}{|r||r|r|r|r|r|r|r|r|r|r|r||c|c|}
\hline
GPC&\kappa^{(0)}&\kappa^{(1)}&\kappa^{(2)}&\kappa^{(3)}&\kappa^{(4)}&\kappa^{(5)}&\kappa^{(6)}&\kappa^{(7)}&\kappa^{(8)}&\kappa^{(9)}&\kappa^{(10)}
& \{r,d+1-s\} & c \\ \hline
 47 & 8 & -3 & -1 & -2 & -2 & -3 & 0 & 0 & -1 & -2 & 0 & \text{$\{$4, 5$\}$} & \text{1/2} \\ \hline
 48 & 6 & -2 & -1 & -3 & 0 & -1 & -2 & 1 & 0 & 0 & -1 & \text{$\{$7$\}$} & \text{3/2} \\ \hline
 49 & 6 & -1 & -2 & -3 & 0 & -2 & -1 & 1 & 0 & 0 & -1 & \text{$\{$7$\}$} & \text{3/2} \\ \hline
 50 & 6 & -3 & -2 & 0 & 0 & -1 & -1 & -2 & 1 & -1 & 0 & & \text{6/5} \\ \hline
 51 & 6 & -3 & -2 & 0 & 0 & -1 & -1 & -1 & 0 & -2 & 1 & & \text{6/5} \\ \hline
 52 & 6 & -3 & -2 & -1 & 0 & 0 & 0 & -1 & -1 & -2 & 1 & \text{$\{$4, 5$\}$} & \text{3/4} \\ \hline
 53 & 6 & -3 & -1 & -2 & 0 & -2 & 0 & 1 & 0 & -1 & -1 & \text{$\{$2, 7$\}$} & \text{3/2} \\ \hline
 54 & 6 & -3 & -1 & -2 & 0 & -2 & 1 & 0 & -1 & 0 & -1 & \text{$\{$4, 5$\}$} & \text{7/12} \\ \hline
 55 & 6 & -3 & 0 & -1 & -2 & -2 & -1 & 1 & 0 & 0 & -1 & \text{$\{$7$\}$} & \text{3/2} \\ \hline
 56 & 6 & -3 & 0 & -2 & -1 & -1 & -2 & 1 & 0 & 0 & -1 & \text{$\{$7$\}$} & \text{3/2} \\ \hline
 57 & 6 & -3 & -2 & -1 & 0 & -2 & 1 & 0 & 0 & -1 & -1 & \text{$\{$2, 7$\}$} & \text{5/4} \\ \hline
 58 & 6 & -3 & -2 & -2 & 1 & 0 & 0 & -1 & -1 & -1 & 0 & \text{$\{$4, 5$\}$} & \text{3/4} \\ \hline
 59 & 6 & -2 & -1 & -1 & -2 & -3 & 0 & 1 & 0 & 0 & -1 & \text{$\{$7$\}$} & \text{3/2} \\ \hline
 60 & 6 & -2 & -1 & -1 & -2 & -3 & 1 & 0 & 0 & -1 & 0 & \text{$\{$4, 5$\}$} & \text{7/12} \\ \hline
 61 & 6 & -3 & -2 & -2 & 1 & -1 & 0 & 0 & 0 & -1 & -1 & \text{$\{$2, 7$\}$} & \text{5/4} \\ \hline
 62 & 6 & -1 & -2 & -2 & -1 & -3 & 0 & 1 & 0 & 0 & -1 & \text{$\{$7$\}$} & \text{3/2} \\ \hline
 63 & 6 & -2 & -1 & -2 & -1 & -3 & 1 & 0 & 0 & 0 & -1 & \text{$\{$4, 5$\}$} & \text{11/20} \\ \hline
 64 & 6 & -1 & -2 & -2 & -1 & -3 & 1 & 0 & 0 & -1 & 0 & \text{$\{$4, 5$\}$} & \text{7/12} \\ \hline
 65 & 8 & -2 & -2 & -3 & -1 & -5 & 2 & 2 & 1 & 0 & 0 & \text{$\{$4, 5$\}$} & 1 \\ \hline
 66 & 8 & -3 & -1 & -2 & -2 & -5 & 2 & 2 & 1 & 0 & 0 & \text{$\{$4, 5$\}$} & 1 \\ \hline
 67 & 8 & -5 & 0 & -1 & -2 & -3 & 2 & 1 & 0 & -2 & 2 & \text{$\{$4, 5$\}$} & 1 \\ \hline
 68 & 8 & -5 & -2 & -3 & 2 & 0 & 1 & 0 & -1 & -2 & 2 & \text{$\{$4, 5$\}$} & \text{5/4} \\ \hline
 69 & 8 & -5 & -3 & 1 & 0 & 0 & -1 & -2 & 2 & -2 & 2 & & \text{12/5} \\ \hline
 70 & 8 & -5 & -3 & -2 & 2 & 1 & 0 & 0 & -1 & -2 & 2 & \text{$\{$4, 5$\}$} & \text{5/4} \\ \hline
 71 & 8 & -5 & -2 & -3 & 2 & -2 & 2 & 1 & 0 & -1 & 0 & \text{$\{$4, 5$\}$} & \text{7/6} \\ \hline
 72 & 8 & -5 & -2 & -2 & 1 & -3 & 2 & 2 & 0 & 0 & -1 & \text{$\{$2, 7$\}$} & \text{11/4} \\ \hline
 73 & 8 & -5 & -2 & -3 & 2 & -2 & 1 & 2 & 0 & 0 & -1 & \text{$\{$2, 7$\}$} & \text{11/4} \\ \hline
 74 & 8 & -5 & -3 & -2 & 2 & -2 & 2 & 1 & 0 & 0 & -1 & \text{$\{$2, 7$\}$} & \text{5/2} \\ \hline
 75 & 4 & -2 & -3 & -1 & 2 & -1 & 2 & 0 & 1 & 1 & 0 & \text{$\{$2$\}$} & 2 \\ \hline
 76 & 4 & -2 & -3 & 0 & 1 & -1 & 2 & -1 & 2 & 1 & 0 & \text{$\{$2$\}$} & 2 \\ \hline
 77 & 4 & -1 & -3 & -1 & 1 & -2 & 2 & 0 & 2 & 1 & 0 & \text{$\{$2, 7$\}$} & 2 \\ \hline
 78 & 4 & -2 & 0 & -3 & 1 & -1 & -1 & 2 & 2 & 1 & 0 & \text{$\{$7$\}$} & \text{3/2} \\ \hline
 79 & 4 & -2 & -1 & -3 & 2 & 0 & -1 & 1 & 2 & 1 & 0 & \text{$\{$2, 7$\}$} & \text{7/4} \\ \hline
 80 & 4 & -1 & -1 & -3 & 1 & -2 & 0 & 2 & 2 & 1 & 0 & \text{$\{$7$\}$} & \text{3/2} \\ \hline
 81 & 4 & -1 & -2 & -3 & 2 & -1 & 0 & 1 & 2 & 1 & 0 & \text{$\{$2, 7$\}$} & \text{7/4} \\ \hline
 82 & 4 & -2 & -3 & -1 & 2 & 0 & 1 & -1 & 2 & 1 & 0 & \text{$\{$2$\}$} & 2 \\ \hline
 83 & 4 & -1 & -3 & -2 & 2 & -1 & 1 & 0 & 2 & 1 & 0 & \text{$\{$2, 7$\}$} & 2 \\ \hline
 84 & 4 & -3 & -1 & 1 & -1 & -2 & 2 & 0 & 2 & 1 & 0 & \text{$\{$2, 7$\}$} & 2 \\ \hline
 85 & 4 & -3 & -1 & -2 & 2 & 1 & -1 & 0 & 2 & 1 & 0 & \text{$\{$2, 7$\}$} & 2 \\ \hline
 86 & 4 & -3 & -2 & 1 & 0 & -1 & 2 & -1 & 2 & 1 & 0 & \text{$\{$2$\}$} & 2 \\ \hline
 87 & 4 & -3 & -2 & -1 & 2 & 1 & 0 & -1 & 2 & 1 & 0 & \text{$\{$2$\}$} & 2 \\ \hline
 88 & 4 & -3 & 1 & -1 & -1 & -2 & 0 & 2 & 2 & 1 & 0 & \text{$\{$7$\}$} & \text{3/2} \\ \hline
 89 & 4 & -3 & 1 & -2 & 0 & -1 & -1 & 2 & 2 & 1 & 0 & \text{$\{$7$\}$} & \text{3/2} \\ \hline
 90 & 4 & -2 & -1 & -3 & 2 & -1 & 0 & 2 & 1 & 1 & 0 & \text{$\{$2, 7$\}$} & \text{7/4} \\ \hline
 91 & 4 & -3 & -1 & -2 & 2 & -1 & 1 & 2 & 0 & 1 & 0 & \text{$\{$2, 7$\}$} & 2 \\ \hline
 92 & 4 & -3 & -1 & -1 & 1 & -2 & 2 & 2 & 0 & 1 & 0 & \text{$\{$2, 7$\}$} & 2 \\ \hline
\end{array}
$
\end{minipage}

\renewcommand{\arraystretch}{1.14}
\begin{minipage}{18.0cm}
$\hspace{1.7cm}
\begin{array}{|r||r|r|r|r|r|r|r|r|r|r|r||c|c|}
\hline
GPC&\kappa^{(0)}&\kappa^{(1)}&\kappa^{(2)}&\kappa^{(3)}&\kappa^{(4)}&\kappa^{(5)}&\kappa^{(6)}&\kappa^{(7)}&\kappa^{(8)}&\kappa^{(9)}&\kappa^{(10)}
& \{r,d+1-s\} & c \\ \hline
 93 & 4 & -3 & -2 & -1 & 2 & -1 & 2 & 1 & 0 & 1 & 0 & \text{$\{$2$\}$} & 2 \\ \hline
 94 & 4 & -2 & -1 & 0 & -1 & -3 & 2 & 1 & 2 & 1 & 0 & \text{$\{$2, 7$\}$} & \text{7/4} \\ \hline
 95 & 4 & -1 & -2 & -1 & 0 & -3 & 2 & 1 & 2 & 1 & 0 & \text{$\{$2, 7$\}$} & \text{7/4} \\ \hline
 96 & 4 & -2 & 0 & -1 & -1 & -3 & 1 & 2 & 2 & 1 & 0 & \text{$\{$7$\}$} & \text{3/2} \\ \hline
 97 & 4 & -1 & -1 & -2 & 0 & -3 & 1 & 2 & 2 & 1 & 0 & \text{$\{$7$\}$} & \text{3/2} \\ \hline
 98 & 4 & -2 & -1 & -1 & 0 & -3 & 2 & 2 & 1 & 1 & 0 & \text{$\{$2, 7$\}$} & \text{7/4} \\ \hline
 99 & 4 & -3 & -2 & 1 & 0 & -1 & 2 & 1 & 0 & -1 & 2 & \text{$\{$2$\}$} & 2 \\ \hline
 100 & 4 & -1 & -2 & -3 & 2 & -1 & 1 & 0 & 2 & 0 & 1 & \text{$\{$4, 5$\}$} & \text{3/4} \\ \hline
 101 & 4 & -2 & -3 & -1 & 2 & 1 & 0 & 0 & 1 & -1 & 2 & \text{$\{$2$\}$} & 2 \\ \hline
 102 & 4 & -2 & -3 & -1 & 2 & 1 & 0 & -1 & 2 & 0 & 1 & \text{$\{$2$\}$} & 2 \\ \hline
 103 & 4 & -2 & -3 & 1 & 0 & -1 & 2 & -1 & 2 & 0 & 1 & \text{$\{$2$\}$} & 2 \\ \hline
 104 & 4 & -2 & -3 & 1 & 0 & 0 & 1 & -1 & 2 & -1 & 2 & \text{$\{$2$\}$} & 2 \\ \hline
 105 & 4 & -2 & -3 & 1 & 0 & -1 & 2 & 0 & 1 & -1 & 2 & \text{$\{$2$\}$} & 2 \\ \hline
 106 & 4 & -2 & -3 & -1 & 2 & 0 & 1 & 1 & 0 & -1 & 2 & \text{$\{$2$\}$} & 2 \\ \hline
 107 & 4 & -2 & -3 & 0 & 1 & -1 & 2 & 1 & 0 & -1 & 2 & \text{$\{$2$\}$} & 2 \\ \hline
 108 & 4 & -2 & -3 & 0 & 1 & 1 & 0 & -1 & 2 & -1 & 2 & \text{$\{$2$\}$} & 2 \\ \hline
 109 & 4 & -3 & -2 & 1 & 0 & 1 & 0 & -1 & 2 & -1 & 2 & \text{$\{$2$\}$} & 2 \\ \hline
 110 & 4 & -3 & -2 & -1 & 2 & 1 & 0 & 1 & 0 & -1 & 2 & \text{$\{$2$\}$} & 2 \\ \hline
 111 & 4 & -2 & -3 & -1 & 2 & -1 & 2 & 1 & 0 & 0 & 1 & \text{$\{$2$\}$} & 2 \\ \hline
 112 & 4 & -2 & -1 & 0 & -1 & -3 & 2 & 2 & 1 & 0 & 1 & \text{$\{$2, 7$\}$} & 2 \\ \hline
 113 & 4 & -3 & -1 & -2 & 2 & -1 & 2 & 1 & 0 & 0 & 1 & \text{$\{$4, 5$\}$} & \text{5/6} \\ \hline
 114 & 4 & -2 & 0 & -1 & -1 & -3 & 2 & 1 & 2 & 0 & 1 & \text{$\{$4, 5$\}$} & \text{3/4} \\ \hline
 115 & 4 & -2 & -1 & -3 & 2 & 0 & -1 & 2 & 1 & 0 & 1 & \text{$\{$2, 7$\}$} & 2 \\ \hline
 116 & 4 & -3 & 0 & 0 & -1 & -2 & 2 & 1 & 1 & -1 & 2 & \text{$\{$4, 5$\}$} & \text{3/4} \\ \hline
 117 & 4 & -3 & -1 & -2 & 2 & 0 & 1 & 1 & 0 & -1 & 2 & \text{$\{$4, 5$\}$} & \text{5/6} \\ \hline
 118 & 4 & -1 & -2 & -1 & 0 & -3 & 2 & 2 & 1 & 0 & 1 & \text{$\{$2, 7$\}$} & 2 \\ \hline
 119 & 4 & -1 & -1 & -2 & 0 & -3 & 2 & 1 & 2 & 0 & 1 & \text{$\{$4, 5$\}$} & \text{3/4} \\ \hline
 120 & 4 & -1 & -2 & -3 & 2 & -1 & 0 & 2 & 1 & 0 & 1 & \text{$\{$2, 7$\}$} & 2 \\ \hline
 121 & 7 & -2 & -3 & -4 & 2 & -2 & 0 & 2 & 1 & 0 & -1 & \text{$\{$2, 7$\}$} & \text{5/2} \\ \hline
 122 & 7 & -2 & -2 & -3 & 0 & -4 & 2 & 1 & 2 & 0 & -1 & \text{$\{$4, 5$\}$} & 1 \\ \hline
 123 & 7 & -2 & -2 & -3 & 0 & -4 & 2 & 2 & 1 & -1 & 0 & \text{$\{$4, 5$\}$} & 1 \\ \hline
 124 & 7 & -4 & -2 & -3 & 2 & -2 & 2 & 1 & 0 & 0 & -1 & \text{$\{$4, 5$\}$} & \text{25/24} \\ \hline
 125 & 7 & -2 & -3 & -2 & 0 & -4 & 2 & 2 & 1 & 0 & -1 & \text{$\{$2, 7$\}$} & \text{5/2} \\ \hline
\end{array}
$
\end{minipage}

\subsection*{Setting$(N,d)=(5,10)$}
\renewcommand{\arraystretch}{1.14}
\begin{minipage}{18.0cm}
$\hspace{1.7cm}
\begin{array}{|r||r|r|r|r|r|r|r|r|r|r|r||c|c|}
\hline
GPC&\kappa^{(0)}&\kappa^{(1)}&\kappa^{(2)}&\kappa^{(3)}&\kappa^{(4)}&\kappa^{(5)}&\kappa^{(6)}&\kappa^{(7)}&\kappa^{(8)}&\kappa^{(9)}&\kappa^{(10)}
& \{r,d+1-s\} & c \\ \hline
 1 & 1 & -1 & 0 & 0 & 0 & 0 & 0 & 0 & 0 & 0 & 0 & \text{$\{$1$\}$} & \text{1/2} \\ \hline
 2 & 4 & 0 & -1 & -1 & -1 & -1 & -1 & -1 & 0 & 0 & 0 & \text{$\{$1, 8$\}$} & \text{1/2} \\ \hline
 3 & 3 & -1 & -1 & -1 & 0 & 0 & 0 & 0 & 0 & 0 & -1 & \text{$\{$3, 10$\}$} & \text{1/2} \\ \hline
 4 & 2 & 0 & -1 & -1 & -1 & 1 & 0 & 0 & 0 & 1 & 1 & \text{$\{$3, 8$\}$} & \text{5/6} \\ \hline
 5 & 2 & -1 & -1 & 0 & 0 & 0 & -1 & 1 & 1 & 1 & 0 & \text{$\{$3, 8$\}$} & \text{5/6} \\ \hline
 6 & 2 & 0 & -1 & -1 & 0 & 0 & -1 & 1 & 0 & 1 & 1 & \text{$\{$3, 8$\}$} & \text{5/6} \\ \hline
 7 & 2 & -1 & -1 & -1 & 0 & 1 & 0 & 1 & 0 & 1 & 0 & \text{$\{$3, 10$\}$} & \text{3/4} \\ \hline
 8 & 2 & -1 & -1 & 0 & -1 & 1 & 0 & 0 & 1 & 1 & 0 & \text{$\{$3, 8$\}$} & \text{5/6} \\ \hline
 9 & 2 & 0 & -1 & 0 & -1 & 0 & -1 & 0 & 1 & 1 & 1 & \text{$\{$1, 8$\}$} & \text{3/4} \\ \hline
 10 & 6 & -2 & -2 & -3 & 1 & 0 & 1 & 0 & -1 & 2 & -1 & \text{$\{$3, 10$\}$} & 2 \\ \hline
 11 & 6 & -2 & -2 & -3 & -1 & 2 & 1 & 0 & 1 & 0 & -1 & \text{$\{$3, 10$\}$} & 2 \\ \hline
 12 & 6 & 0 & -1 & -2 & -3 & 0 & -1 & -2 & 2 & 1 & 1 & \text{$\{$1, 8$\}$} & 2 \\ \hline
 13 & 6 & -2 & -2 & -3 & 1 & 0 & -1 & 2 & 1 & 0 & -1 & \text{$\{$3, 10$\}$} & 2 \\ \hline
 14 & 6 & 0 & -3 & 0 & -1 & -2 & -1 & -2 & 2 & 1 & 1 & \text{$\{$1, 8$\}$} & 2 \\ \hline
 15 & 6 & 0 & -3 & 0 & -1 & -1 & -2 & -2 & 1 & 2 & 1 & & \text{17/9} \\ \hline
 16 & 6 & 0 & -1 & -2 & -1 & -2 & -3 & 0 & 2 & 1 & 1 & \text{$\{$1, 8$\}$} & 2 \\ \hline
 17 & 6 & -2 & -3 & -2 & -1 & 2 & 1 & 1 & 0 & 0 & -1 & \text{$\{$5, 6$\}$} & 1 \\ \hline
 18 & 6 & 0 & -1 & -1 & -2 & -2 & -3 & 0 & 1 & 2 & 1 & \text{$\{$5, 6$\}$} & 1 \\ \hline
 19 & 6 & -2 & -3 & -2 & 1 & 1 & 0 & 0 & -1 & 2 & -1 & & \text{17/9} \\ \hline
 20 & 7 & -1 & -2 & -1 & -3 & 0 & -3 & 0 & 2 & 2 & 1 & \text{$\{$1, 8$\}$} & \text{7/4} \\ \hline
 21 & 7 & -1 & -3 & 0 & -3 & 0 & -2 & -1 & 2 & 2 & 1 & \text{$\{$1, 8$\}$} & \text{7/4} \\ \hline
 22 & 7 & -2 & -3 & 0 & -3 & 1 & -1 & -1 & 2 & 2 & 0 & \text{$\{$3, 8$\}$} & \text{11/6} \\ \hline
 23 & 7 & -2 & -3 & -1 & -3 & 2 & -1 & 0 & 2 & 1 & 0 & \text{$\{$3, 8$\}$} & 2 \\ \hline
 24 & 7 & -1 & -3 & 0 & -2 & -1 & -3 & 0 & 2 & 2 & 1 & \text{$\{$1, 8$\}$} & \text{7/4} \\ \hline
 25 & 7 & -2 & -3 & -3 & -1 & 2 & 0 & 1 & -1 & 2 & 0 & \text{$\{$3, 10$\}$} & \text{7/4} \\ \hline
 26 & 7 & -2 & -3 & -3 & -1 & 2 & -1 & 2 & 0 & 1 & 0 & \text{$\{$3, 10$\}$} & \text{7/4} \\ \hline
 27 & 7 & -2 & -3 & -1 & -3 & 2 & 0 & -1 & 1 & 2 & 0 & \text{$\{$3, 8$\}$} & 2 \\ \hline
 28 & 7 & -1 & -2 & -3 & 0 & -1 & -3 & 2 & 0 & 1 & 2 & \text{$\{$3, 8$\}$} & \text{13/6} \\ \hline
 29 & 7 & -1 & -3 & -2 & 0 & -1 & -3 & 2 & 0 & 2 & 1 & \text{$\{$3, 8$\}$} & 2 \\ \hline
 30 & 7 & -1 & -2 & -3 & -1 & 0 & -3 & 2 & 0 & 2 & 1 & \text{$\{$3, 8$\}$} & 2 \\ \hline
 31 & 7 & -2 & -3 & 0 & -1 & -1 & -3 & 1 & 2 & 2 & 0 & \text{$\{$3, 8$\}$} & \text{11/6} \\ \hline
 32 & 7 & -2 & -3 & -1 & -1 & 0 & -3 & 2 & 2 & 1 & 0 & \text{$\{$3, 8$\}$} & 2 \\ \hline
 33 & 7 & -2 & -3 & -3 & 0 & 1 & -1 & 2 & -1 & 2 & 0 & \text{$\{$3, 10$\}$} & \text{7/4} \\ \hline
 34 & 7 & -1 & -3 & -3 & 0 & 0 & -2 & 2 & -1 & 2 & 1 & \text{$\{$3, 8$\}$} & \text{11/6} \\ \hline
 35 & 7 & -3 & -2 & -1 & -3 & 2 & 0 & -1 & 2 & 1 & 0 & \text{$\{$3, 8$\}$} & \text{13/6} \\ \hline
 36 & 7 & -1 & -3 & -3 & -2 & 2 & 0 & 0 & -1 & 2 & 1 & \text{$\{$3, 8$\}$} & \text{11/6} \\ \hline
 37 & 7 & -1 & -2 & -3 & -3 & 2 & -1 & 0 & 0 & 2 & 1 & \text{$\{$3, 8$\}$} & 2 \\ \hline
 38 & 7 & -2 & -3 & -1 & 0 & -1 & -3 & 2 & 1 & 2 & 0 & \text{$\{$3, 8$\}$} & 2 \\ \hline
 39 & 7 & -1 & -3 & -2 & -3 & 2 & 0 & -1 & 0 & 2 & 1 & \text{$\{$3, 8$\}$} & 2 \\ \hline
 40 & 7 & -3 & -2 & -3 & 1 & 0 & -1 & 2 & -1 & 2 & 0 & \text{$\{$3, 10$\}$} & 2 \\ \hline
 41 & 7 & -1 & -3 & 0 & -3 & 0 & -1 & -2 & 2 & 1 & 2 & \text{$\{$1, 8$\}$} & 2 \\ \hline
 42 & 7 & -3 & -2 & 0 & -1 & -1 & -3 & 2 & 1 & 2 & 0 & \text{$\{$5, 6$\}$} & \text{19/20} \\ \hline
 43 & 7 & -1 & -3 & -2 & -3 & 2 & 0 & 0 & -1 & 1 & 2 & \text{$\{$5, 6$\}$} & \text{19/20} \\ \hline
 44 & 7 & -1 & -2 & -3 & -3 & 2 & 0 & -1 & 0 & 1 & 2 & \text{$\{$3, 8$\}$} & \text{13/6} \\ \hline
 45 & 7 & -3 & -2 & -1 & 0 & -1 & -3 & 2 & 2 & 1 & 0 & \text{$\{$3, 8$\}$} & \text{13/6} \\ \hline
 46 & 7 & -3 & -1 & -2 & -3 & 2 & -1 & 1 & 0 & 2 & 0 & \text{$\{$5, 6$\}$} & \text{19/20} \\ \hline
\end{array}
$
\end{minipage}

\renewcommand{\arraystretch}{1.14}
\begin{minipage}{18.0cm}
$\hspace{1.7cm}
\begin{array}{|r||r|r|r|r|r|r|r|r|r|r|r||c|c|}
\hline
GPC&\kappa^{(0)}&\kappa^{(1)}&\kappa^{(2)}&\kappa^{(3)}&\kappa^{(4)}&\kappa^{(5)}&\kappa^{(6)}&\kappa^{(7)}&\kappa^{(8)}&\kappa^{(9)}&\kappa^{(10)}
& \{r,d+1-s\} & c \\ \hline
 47 & 7 & -1 & -3 & -1 & -2 & 0 & -3 & 2 & 1 & 0 & 2 & \text{$\{$5, 6$\}$} & \text{19/20} \\ \hline
 48 & 7 & -3 & -1 & -2 & -3 & 2 & -1 & 0 & 2 & 1 & 0 & \text{$\{$3, 8$\}$} & 2 \\ \hline
 49 & 7 & -1 & -2 & -3 & -1 & 0 & -3 & 2 & 1 & 0 & 2 & \text{$\{$3, 8$\}$} & 2 \\ \hline
 50 & 7 & -1 & -3 & -2 & -3 & 2 & -1 & 1 & 0 & 0 & 2 & \text{$\{$5, 6$\}$} & \text{19/20} \\ \hline
 51 & 7 & -3 & -1 & -1 & -2 & 0 & -3 & 2 & 1 & 2 & 0 & \text{$\{$5, 6$\}$} & \text{19/20} \\ \hline
 52 & 7 & -3 & -1 & -2 & -1 & 0 & -3 & 2 & 2 & 1 & 0 & \text{$\{$3, 8$\}$} & 2 \\ \hline
 53 & 7 & -1 & -2 & -3 & -3 & 2 & -1 & 0 & 1 & 0 & 2 & \text{$\{$3, 8$\}$} & 2 \\ \hline
 54 & 7 & -3 & -2 & -3 & -1 & 2 & 1 & 0 & -1 & 2 & 0 & \text{$\{$3, 10$\}$} & 2 \\ \hline
 55 & 7 & -1 & -3 & 0 & -1 & -2 & -3 & 0 & 2 & 1 & 2 & \text{$\{$1, 8$\}$} & 2 \\ \hline
 56 & 7 & -1 & -1 & -2 & -3 & 0 & -3 & 0 & 2 & 1 & 2 & \text{$\{$1, 8$\}$} & 2 \\ \hline
 57 & 7 & -3 & -2 & -3 & -1 & 2 & -1 & 2 & 1 & 0 & 0 & \text{$\{$3, 10$\}$} & 2 \\ \hline
 58 & 7 & -1 & -1 & -2 & -3 & 0 & -3 & 1 & 0 & 2 & 2 & \text{$\{$5, 6$\}$} & \text{31/30} \\ \hline
 59 & 7 & -3 & -3 & -1 & -2 & 2 & -1 & 2 & 1 & 0 & 0 & \text{$\{$5, 6$\}$} & \text{31/30} \\ \hline
 60 & 9 & -3 & -4 & -5 & -1 & 4 & 3 & 2 & 1 & 0 & -2 & \text{$\{$5, 6$\}$} & 2 \\ \hline
 61 & 9 & -3 & -4 & -5 & 3 & 2 & 1 & 0 & -1 & 4 & -2 & & \text{11/3} \\ \hline
 62 & 9 & 1 & -5 & 0 & -1 & -2 & -3 & -4 & 4 & 3 & 2 & & \text{11/3} \\ \hline
 63 & 9 & 1 & -1 & -2 & -3 & -4 & -5 & 0 & 4 & 3 & 2 & \text{$\{$5, 6$\}$} & 2 \\ \hline
 64 & 4 & -1 & -2 & 0 & -2 & 1 & 0 & -1 & 2 & 1 & 0 & \text{$\{$3, 8$\}$} & \text{7/4} \\ \hline
 65 & 4 & 0 & -1 & -2 & 0 & -1 & -2 & 1 & 0 & 2 & 1 & \text{$\{$3, 8$\}$} & \text{7/4} \\ \hline
 66 & 4 & -2 & -2 & -1 & 1 & 1 & 0 & 0 & -1 & 2 & 0 & & \text{3/2} \\ \hline
 67 & 4 & 0 & -1 & -2 & -2 & 1 & 0 & 0 & -1 & 1 & 2 & \text{$\{$5, 6$\}$} & \text{7/10} \\ \hline
 68 & 4 & -2 & 0 & -1 & -1 & 0 & -2 & 2 & 1 & 1 & 0 & \text{$\{$5, 6$\}$} & \text{13/20} \\ \hline
 69 & 4 & 0 & -1 & -2 & -2 & 1 & 0 & -1 & 0 & 2 & 1 & \text{$\{$3, 8$\}$} & \text{7/4} \\ \hline
 70 & 4 & 0 & -2 & -1 & -2 & 1 & 0 & 0 & -1 & 2 & 1 & \text{$\{$5, 6$\}$} & \text{7/10} \\ \hline
 71 & 4 & -2 & -1 & 0 & 0 & -1 & -2 & 2 & 1 & 1 & 0 & \text{$\{$5, 6$\}$} & \text{11/16} \\ \hline
 72 & 4 & -2 & -2 & -1 & -1 & 2 & 1 & 1 & 0 & 0 & 0 & \text{$\{$5, 6$\}$} & \text{4/5} \\ \hline
 73 & 4 & 0 & -1 & -1 & -2 & 0 & -2 & 1 & 0 & 2 & 1 & \text{$\{$5, 6$\}$} & \text{7/10} \\ \hline
 74 & 4 & -1 & -2 & 0 & -1 & 0 & -2 & 2 & 1 & 1 & 0 & \text{$\{$5, 6$\}$} & \text{13/20} \\ \hline
 75 & 4 & -1 & -2 & 0 & 0 & -1 & -2 & 1 & 2 & 1 & 0 & \text{$\{$3, 8$\}$} & \text{7/4} \\ \hline
 76 & 4 & 0 & -2 & 0 & -1 & -1 & -2 & 0 & 2 & 1 & 1 & \text{$\{$1, 8$\}$} & \text{3/2} \\ \hline
 77 & 4 & 0 & -1 & -1 & -2 & 0 & -2 & 0 & 2 & 1 & 1 & \text{$\{$1, 8$\}$} & \text{3/2} \\ \hline
 78 & 4 & 0 & -2 & 0 & -2 & 0 & -1 & -1 & 2 & 1 & 1 & \text{$\{$1, 8$\}$} & \text{3/2} \\ \hline
 79 & 2 & 0 & -2 & 1 & 0 & 0 & -1 & -1 & 1 & 2 & 2 & & \text{3/2} \\ \hline
 80 & 2 & 0 & -1 & -2 & 1 & 0 & -1 & 2 & 0 & 2 & 1 & \text{$\{$3, 8$\}$} & \text{7/4} \\ \hline
 81 & 2 & -1 & -2 & 0 & -1 & 2 & 1 & 0 & 2 & 1 & 0 & \text{$\{$3, 8$\}$} & \text{7/4} \\ \hline
 82 & 2 & -2 & -1 & 1 & 0 & 0 & -1 & 2 & 2 & 1 & 0 & \text{$\{$5, 6$\}$} & \text{7/10} \\ \hline
 83 & 2 & 0 & -1 & -1 & -2 & 2 & 0 & 1 & 1 & 0 & 2 & \text{$\{$5, 6$\}$} & \text{13/20} \\ \hline
 84 & 2 & -1 & -2 & 0 & 1 & 0 & -1 & 2 & 2 & 1 & 0 & \text{$\{$3, 8$\}$} & \text{7/4} \\ \hline
 85 & 2 & -1 & -2 & 1 & 0 & 0 & -1 & 2 & 1 & 2 & 0 & \text{$\{$5, 6$\}$} & \text{7/10} \\ \hline
 86 & 2 & 0 & -1 & -1 & -2 & 2 & 1 & 0 & 0 & 1 & 2 & \text{$\{$5, 6$\}$} & \text{11/16} \\ \hline
 87 & 2 & -1 & -2 & 0 & -1 & 2 & 0 & 2 & 1 & 1 & 0 & \text{$\{$5, 6$\}$} & \text{7/10} \\ \hline
 88 & 2 & 0 & -1 & -1 & -2 & 2 & 0 & 1 & 0 & 2 & 1 & \text{$\{$5, 6$\}$} & \text{13/20} \\ \hline
 89 & 2 & 0 & -1 & -2 & -1 & 2 & 1 & 0 & 0 & 2 & 1 & \text{$\{$3, 8$\}$} & \text{7/4} \\ \hline
 90 & 2 & -1 & -1 & -2 & 1 & 1 & 0 & 2 & 0 & 2 & 0 & \text{$\{$3, 10$\}$} & \text{3/2} \\ \hline
 91 & 2 & 0 & 0 & 0 & -1 & -1 & -2 & 1 & 1 & 2 & 2 & \text{$\{$5, 6$\}$} & \text{4/5} \\ \hline
 92 & 2 & -1 & -1 & -2 & 0 & 2 & 0 & 2 & 1 & 1 & 0 & \text{$\{$3, 10$\}$} & \text{3/2} \\ \hline
\end{array}
$
\end{minipage}

\renewcommand{\arraystretch}{1.14}
\begin{minipage}{18.0cm}
$\hspace{1.7cm}
\begin{array}{|r||r|r|r|r|r|r|r|r|r|r|r||c|c|}
\hline
GPC&\kappa^{(0)}&\kappa^{(1)}&\kappa^{(2)}&\kappa^{(3)}&\kappa^{(4)}&\kappa^{(5)}&\kappa^{(6)}&\kappa^{(7)}&\kappa^{(8)}&\kappa^{(9)}&\kappa^{(10)}
& \{r,d+1-s\} & c \\ \hline
 93 & 2 & -1 & -1 & -2 & 0 & 2 & 1 & 1 & 0 & 2 & 0 & \text{$\{$3, 10$\}$} & \text{3/2} \\ \hline
 94 & 3 & 0 & -1 & -1 & -1 & 0 & -1 & 0 & 0 & 0 & 1 & \text{$\{$3, 8$\}$} & \text{2/3} \\ \hline
 95 & 3 & -1 & 0 & -1 & -1 & 0 & -1 & 0 & 0 & 1 & 0 & \text{$\{$1, 8$\}$} & 1 \\ \hline
 96 & 3 & -1 & -1 & 0 & 0 & -1 & -1 & 0 & 0 & 1 & 0 & \text{$\{$1, 8$\}$, $\{$2, 9$\}$} & \text{1, 2} \\ \hline
 97 & 3 & -1 & -1 & -1 & 0 & 0 & -1 & 1 & 0 & 0 & 0 & \text{$\{$3, 10$\}$} & \text{1/2} \\ \hline
 98 & 3 & -1 & -1 & 0 & -1 & 0 & 0 & -1 & 0 & 1 & 0 & \text{$\{$1, 8$\}$, $\{$2, 9$\}$} & \text{1, 2} \\ \hline
 99 & 3 & -1 & -1 & 0 & -1 & 0 & -1 & 0 & 1 & 0 & 0 & \text{$\{$1, 8$\}$} & 1 \\ \hline
 100 & 3 & -1 & -1 & -1 & -1 & 1 & 0 & 0 & 0 & 0 & 0 & \text{$\{$3, 10$\}$} & \text{1/2} \\ \hline
 101 & 3 & -1 & -1 & -1 & 0 & 0 & 0 & 0 & -1 & 1 & 0 & \text{$\{$2, 9$\}$, $\{$3, 10$\}$} & \text{2, 1/2} \\ \hline
 102 & 5 & -1 & -2 & -1 & -1 & 0 & -1 & 0 & 0 & -1 & 0 & \text{$\{$3, 10$\}$} & 1 \\ \hline
 103 & 5 & -2 & -1 & -1 & -1 & 0 & -1 & 0 & 0 & 0 & -1 & \text{$\{$3, 8$\}$} & \text{2/3} \\ \hline
 104 & 5 & -1 & -2 & -1 & -1 & 0 & 0 & -1 & -1 & 0 & 0 & \text{$\{$2, 9$\}$, $\{$3, 10$\}$} & \text{2, 1} \\ \hline
 105 & 5 & -1 & -2 & -1 & 0 & -1 & -1 & 0 & -1 & 0 & 0 & \text{$\{$2, 9$\}$, $\{$3, 10$\}$} & \text{2, 1} \\ \hline
 106 & 5 & -1 & -1 & -1 & -1 & -1 & -2 & 0 & 0 & 0 & 0 & \text{$\{$1, 8$\}$} & \text{1/2} \\ \hline
 107 & 5 & -1 & -1 & -2 & -1 & 0 & -1 & 0 & -1 & 0 & 0 & \text{$\{$3, 10$\}$} & 1 \\ \hline
 108 & 5 & -1 & -2 & 0 & -1 & -1 & -1 & -1 & 0 & 0 & 0 & \text{$\{$1, 8$\}$, $\{$2, 9$\}$} & \text{1/2, 2} \\ \hline
 109 & 5 & -1 & -1 & -1 & -2 & 0 & -1 & -1 & 0 & 0 & 0 & \text{$\{$1, 8$\}$} & \text{1/2} \\ \hline
 110 & 6 & 0 & -2 & -1 & -2 & -1 & -3 & 0 & 2 & 2 & 1 & \text{$\{$1, 8$\}$} & \text{7/4} \\ \hline
 111 & 6 & 0 & -3 & 0 & -2 & -1 & -2 & -1 & 2 & 2 & 1 & \text{$\{$1, 8$\}$} & \text{7/4} \\ \hline
 112 & 6 & 0 & -2 & -1 & -3 & 0 & -2 & -1 & 2 & 2 & 1 & \text{$\{$1, 8$\}$} & \text{7/4} \\ \hline
 113 & 6 & 0 & -3 & 0 & -1 & -2 & -1 & -2 & 1 & 2 & 2 & \text{$\{$1, 8$\}$} & 2 \\ \hline
 114 & 6 & 0 & -1 & -2 & -2 & -1 & -3 & 1 & 0 & 2 & 2 & \text{$\{$5, 6$\}$} & \text{31/30} \\ \hline
 115 & 6 & 0 & -2 & -1 & -1 & -2 & -3 & 0 & 2 & 1 & 2 & \text{$\{$1, 8$\}$} & 2 \\ \hline
 116 & 6 & 0 & -2 & -1 & -3 & 0 & -1 & -2 & 2 & 1 & 2 & \text{$\{$1, 8$\}$} & 2 \\ \hline
 117 & 6 & 0 & -1 & -2 & -1 & -2 & -3 & 0 & 1 & 2 & 2 & \text{$\{$1, 8$\}$} & 2 \\ \hline
 118 & 6 & 0 & -1 & -2 & -3 & 0 & -1 & -2 & 1 & 2 & 2 & \text{$\{$1, 8$\}$} & 2 \\ \hline
 119 & 6 & 0 & -1 & -2 & -3 & 0 & -2 & -1 & 2 & 1 & 2 & \text{$\{$1, 8$\}$} & 2 \\ \hline
 120 & 6 & 0 & -3 & 0 & -2 & -1 & -1 & -2 & 2 & 1 & 2 & \text{$\{$1, 8$\}$} & 2 \\ \hline
 121 & 6 & 0 & -3 & 0 & -1 & -2 & -2 & -1 & 2 & 1 & 2 & \text{$\{$1, 8$\}$} & 2 \\ \hline
 122 & 6 & 0 & -2 & -2 & -3 & 1 & 0 & -1 & -1 & 2 & 2 & \text{$\{$5, 6$\}$} & \text{19/20} \\ \hline
 123 & 6 & 0 & -1 & -2 & -2 & -1 & -3 & 0 & 2 & 1 & 2 & \text{$\{$1, 8$\}$} & 2 \\ \hline
 124 & 6 & -2 & -2 & 0 & -1 & -1 & -3 & 2 & 2 & 1 & 0 & \text{$\{$5, 6$\}$} & \text{17/20} \\ \hline
 125 & 7 & -3 & -2 & -3 & 1 & 0 & 0 & 1 & -1 & 2 & -1 & \text{$\{$3, 10$\}$} & 2 \\ \hline
 126 & 7 & -3 & -2 & -3 & 0 & 1 & 1 & 0 & -1 & 2 & -1 & \text{$\{$3, 10$\}$} & 2 \\ \hline
 127 & 7 & -3 & -3 & -2 & 1 & 0 & 1 & 0 & -1 & 2 & -1 & \text{$\{$3, 10$\}$} & 2 \\ \hline
 128 & 7 & -3 & -3 & -1 & -2 & 2 & 0 & 1 & 1 & 0 & -1 & \text{$\{$5, 6$\}$} & \text{31/30} \\ \hline
 129 & 7 & -3 & -2 & -3 & -1 & 2 & 1 & 0 & 0 & 1 & -1 & \text{$\{$3, 10$\}$} & 2 \\ \hline
 130 & 7 & -3 & -2 & -3 & 1 & 0 & -1 & 2 & 0 & 1 & -1 & \text{$\{$3, 10$\}$} & 2 \\ \hline
 131 & 7 & -3 & -3 & -2 & -1 & 2 & 1 & 0 & 1 & 0 & -1 & \text{$\{$3, 10$\}$} & 2 \\ \hline
 132 & 7 & -3 & -3 & -2 & 1 & 0 & -1 & 2 & 1 & 0 & -1 & \text{$\{$3, 10$\}$} & 2 \\ \hline
 133 & 7 & -3 & -2 & -3 & 0 & 1 & -1 & 2 & 1 & 0 & -1 & \text{$\{$3, 10$\}$} & 2 \\ \hline
 134 & 7 & -3 & -3 & 0 & 0 & -1 & -2 & 2 & 1 & 1 & -1 & \text{$\{$5, 6$\}$} & \text{19/20} \\ \hline
 135 & 7 & -3 & -2 & -3 & -1 & 2 & 0 & 1 & 1 & 0 & -1 & \text{$\{$3, 10$\}$} & 2 \\ \hline
 136 & 7 & -2 & -3 & -3 & 0 & 1 & -1 & 2 & 0 & 1 & -1 & \text{$\{$3, 10$\}$} & \text{7/4} \\ \hline
 137 & 7 & -2 & -3 & -3 & -1 & 2 & 0 & 1 & 0 & 1 & -1 & \text{$\{$3, 10$\}$} & \text{7/4} \\ \hline
 138 & 7 & -2 & -3 & -3 & 0 & 1 & 0 & 1 & -1 & 2 & -1 & \text{$\{$3, 10$\}$} & \text{7/4} \\ \hline
\end{array}
$
\end{minipage}

\renewcommand{\arraystretch}{1.14}
\begin{minipage}{18.0cm}
$\hspace{1.7cm}
\begin{array}{|r||r|r|r|r|r|r|r|r|r|r|r||c|c|}
\hline
GPC&\kappa^{(0)}&\kappa^{(1)}&\kappa^{(2)}&\kappa^{(3)}&\kappa^{(4)}&\kappa^{(5)}&\kappa^{(6)}&\kappa^{(7)}&\kappa^{(8)}&\kappa^{(9)}&\kappa^{(10)}
& \{r,d+1-s\} & c \\ \hline
 139 & 7 & -1 & -2 & -3 & -3 & 2 & 0 & 0 & -1 & 1 & 1 & \text{$\{$5, 6$\}$} & \text{17/20} \\ \hline
 140 & 12 & -2 & -3 & -4 & -5 & 2 & 0 & -1 & -2 & 1 & 0 & \text{$\{$5, 6$\}$} & \text{9/8} \\ \hline
 141 & 12 & -4 & -5 & 0 & -1 & -2 & -3 & 2 & 1 & 0 & -2 & \text{$\{$5, 6$\}$} & \text{6/5} \\ \hline
 142 & 12 & -4 & -5 & -3 & 1 & 0 & 0 & -1 & -2 & 2 & -2 & & \text{8/3} \\ \hline
 143 & 12 & -4 & -5 & -3 & -2 & 2 & 1 & 0 & 0 & -1 & -2 & \text{$\{$5, 6$\}$} & \text{43/30} \\ \hline
 144 & 12 & -4 & -5 & -2 & -3 & 2 & 0 & 1 & 0 & -1 & -2 & \text{$\{$5, 6$\}$} & \text{7/5} \\ \hline
 145 & 8 & -2 & -3 & 0 & -1 & -2 & -4 & 3 & 2 & 1 & 0 & \text{$\{$5, 6$\}$} & \text{9/8} \\ \hline
 146 & 8 & 0 & -2 & -3 & -4 & 1 & 0 & -1 & -2 & 3 & 2 & \text{$\{$5, 6$\}$} & \text{6/5} \\ \hline
 147 & 8 & 0 & -1 & -2 & -3 & -2 & -4 & 1 & 0 & 3 & 2 & \text{$\{$5, 6$\}$} & \text{7/5} \\ \hline
 148 & 8 & 0 & -4 & 0 & -1 & -2 & -2 & -3 & 1 & 3 & 2 & & \text{8/3} \\ \hline
 149 & 8 & 0 & -1 & -2 & -2 & -3 & -4 & 0 & 1 & 3 & 2 & \text{$\{$5, 6$\}$} & \text{43/30} \\ \hline
 150 & 4 & -2 & -2 & -1 & 0 & 1 & 1 & 1 & 0 & 0 & -1 & \text{$\{$5, 6$\}$} & \text{4/5} \\ \hline
 151 & 4 & -2 & -2 & -1 & 1 & 1 & 0 & 0 & 0 & 1 & -1 & & \text{13/9} \\ \hline
 152 & 6 & 0 & -2 & -1 & -1 & -1 & -2 & -2 & 0 & 1 & 1 & & \text{13/9} \\ \hline
 153 & 6 & 0 & -1 & -1 & -2 & -2 & -2 & -1 & 0 & 1 & 1 & \text{$\{$5, 6$\}$} & \text{4/5} \\ \hline
 154 & 15 & -1 & -7 & 0 & -2 & -3 & -4 & -5 & 3 & 2 & 1 & & \text{11/3} \\ \hline
 155 & 15 & -1 & -2 & -3 & -4 & -5 & -7 & 0 & 3 & 2 & 1 & \text{$\{$5, 6$\}$} & 2 \\ \hline
 156 & 9 & -3 & -4 & -5 & 3 & 2 & 1 & 0 & -2 & 5 & -1 & & \text{11/3} \\ \hline
 157 & 9 & -3 & -4 & -5 & -2 & 5 & 3 & 2 & 1 & 0 & -1 & \text{$\{$5, 6$\}$} & 2 \\ \hline
 158 & 8 & 0 & -1 & -2 & -2 & -3 & -4 & 0 & 2 & 2 & 1 & \text{$\{$5, 6$\}$} & \text{4/3} \\ \hline
 159 & 8 & 0 & -4 & 0 & -1 & -2 & -2 & -3 & 2 & 2 & 1 & & \text{22/9} \\ \hline
 160 & 11 & -3 & -4 & -4 & -2 & 2 & 1 & 0 & 0 & -1 & -2 & \text{$\{$5, 6$\}$} & \text{4/3} \\ \hline
 161 & 11 & -3 & -4 & -4 & 1 & 0 & 0 & -1 & -2 & 2 & -2 & & \text{22/9} \\ \hline
\end{array}
$
\end{minipage}

\end{document}